\def\mode{1}  
\if 0\mode
  \documentclass[acmsmall,review]{acmart}
  \definecolor{revisionColor}{rgb}{1, 0.35, 0.0}
  \newcommand{\revision}[1]{{\color{revisionColor}#1}}
\else
  \documentclass[acmsmall]{acmart}
  \newcommand{\revision}[1]{{\color{black}#1}}
\fi

\usepackage{color}
\usepackage{float}
\usepackage{subcaption}
\usepackage{xspace} 
\usepackage{graphicx}
\usepackage{multirow}
\usepackage{array}
\usepackage{enumitem}
\usepackage{ctable} 
\usepackage{longtable}

\AtBeginDocument{%
  \providecommand\BibTeX{{%
    \normalfont B\kern-0.5em{\scshape i\kern-0.25em b}\kern-0.8em\TeX}}}

\setcopyright{rightsretained}
\acmJournal{PACMHCI}
\acmYear{2021} 
\acmVolume{5} 
\acmNumber{CSCW1} 
\acmArticle{166} 
\acmMonth{4} 
\acmPrice{}
\acmDOI{10.1145/3449240}

\received{June 2020}
\received[revised]{October 2020}
\received[accepted]{December 2020} 

\newcommand{\bolderandunderline}[1]{\textbf{\underline{#1}}}
\newcommand{\dottedExp}[1]{\textbullet~\textit{#1}}
\newcommand{\dottedPriorWork}[1]{\textbullet~#1}

\newcommand{\systemname}{Strata\xspace} 
\newcommand{\systemnameInFull}{\bolderandunderline{S}idebar \bolderandunderline{T}owards \bolderandunderline{R}euse and to \bolderandunderline{A}ssess \bolderandunderline{T}rustworthiness and \bolderandunderline{A}pplicability\xspace}
\newcommand{\stackoverflow}{Stack Overflow\xspace}
\newcommand{\unakite}{Unakite\xspace}
\newcommand{\javascript}{JavaScript\xspace}

\makeatletter
\newcommand*{\bigcdot}{}
\DeclareRobustCommand*{\bigcdot}{%
  \mathbin{\mathpalette\bigcdot@{}}%
}
\newcommand*{\bigcdot@scalefactor}{.5}
\newcommand*{\bigcdot@widthfactor}{1.15}
\newcommand*{\bigcdot@}[2]{%
  \sbox0{$#1\vcenter{}$}
  \sbox2{$#1\cdot\m@th$}%
  \hbox to \bigcdot@widthfactor\wd2{%
    \hfil
    \raise\ht0\hbox{%
      \scalebox{\bigcdot@scalefactor}{%
        \lower\ht0\hbox{$#1\bullet\m@th$}%
      }%
    }%
    \hfil
  }%
}
\makeatother

\begin{document}






\title{To Reuse or Not To Reuse? A Framework and System for Evaluating Summarized Knowledge}






\author{Michael Xieyang Liu}
\affiliation{%
  \institution{Human-Computer Interaction Institute, Carnegie Mellon University}
  \city{Pittsburgh, PA}
  \country{USA}
  }
\email{xieyangl@cs.cmu.edu}

\author{Aniket Kittur}
\affiliation{%
  \institution{Human-Computer Interaction Institute, Carnegie Mellon University}
  \city{Pittsburgh, PA}
  \country{USA}
  }
\email{nkittur@cs.cmu.edu}

\author{Brad A. Myers}
\affiliation{%
  \institution{Human-Computer Interaction Institute, Carnegie Mellon University}
  \city{Pittsburgh, PA}
  \country{USA}
  }
\email{bam@cs.cmu.edu}

\renewcommand{\shortauthors}{Michael Xieyang Liu, Aniket Kittur, \& Brad A. Myers}

\begin{abstract}
As the amount of information online continues to grow, a correspondingly important opportunity is for individuals to reuse knowledge which has been summarized by others rather than starting from scratch. However, appropriate reuse requires judging the relevance, trustworthiness, and thoroughness of others' knowledge in relation to an individual's goals and context. In this work, we explore augmenting judgements of the appropriateness of reusing knowledge in the domain of programming, specifically of reusing artifacts that result from other developers' searching and decision making. Through an analysis of prior research on sensemaking and trust, along with new interviews with developers, we synthesized a framework for reuse judgements. The interviews also validated that developers express a desire for help with judging whether to reuse an existing decision. From this framework, we developed a set of techniques for capturing the initial decision maker's behavior and visualizing signals calculated based on the behavior, to facilitate subsequent consumers' reuse decisions, instantiated in a prototype system called \systemname. Results of a user study suggest that the system significantly improves the accuracy, depth, and speed of reusing decisions. These results have implications for systems involving user-generated content in which other users need to evaluate the relevance and trustworthiness of that content.
\end{abstract}



\maketitle

%
%
%
%
%
%
\section{Introduction}
\label{sec:intro}

Information and knowledge reuse has become a highly consistent paradigm across a wide range of fields and disciplines to advance their respective frontiers, such as reusing previous engineering best practices on future generations of products \cite{baxter_framework_2008,baxter_engineering_2007}, taking advantage of schemas and results from previous sensemaking episodes to create new representations and understandings of the world \cite{kittur_standing_2014,fisher_distributed_2012,paul_cosense:_2009,morris_searchtogether:_2007}, and plugging in previously written and well-maintained \revision{design patterns and code snippets} to build novel software features and functionalities \cite{krueger_software_1992,frakes_software_1986,frakes_software_1996,noauthor_npm_nodate,noauthor_pip_2020}. Reusing proven information and knowledge promises the benefits of potentially reduced workload and development cycles \cite{krueger_software_1992,baxter_engineering_2007}, improved quality and performance \cite{haefliger_code_2007,frakes_software_1986,watson_multi-theoretical_2006}, and more time for creation and innovation \cite{majchrzak_knowledge_2004,markus_toward_2001,watson_multi-theoretical_2006,hurley_innovation_1998}.

\begin{figure}[t]
\centering
	\includegraphics[width=\linewidth]{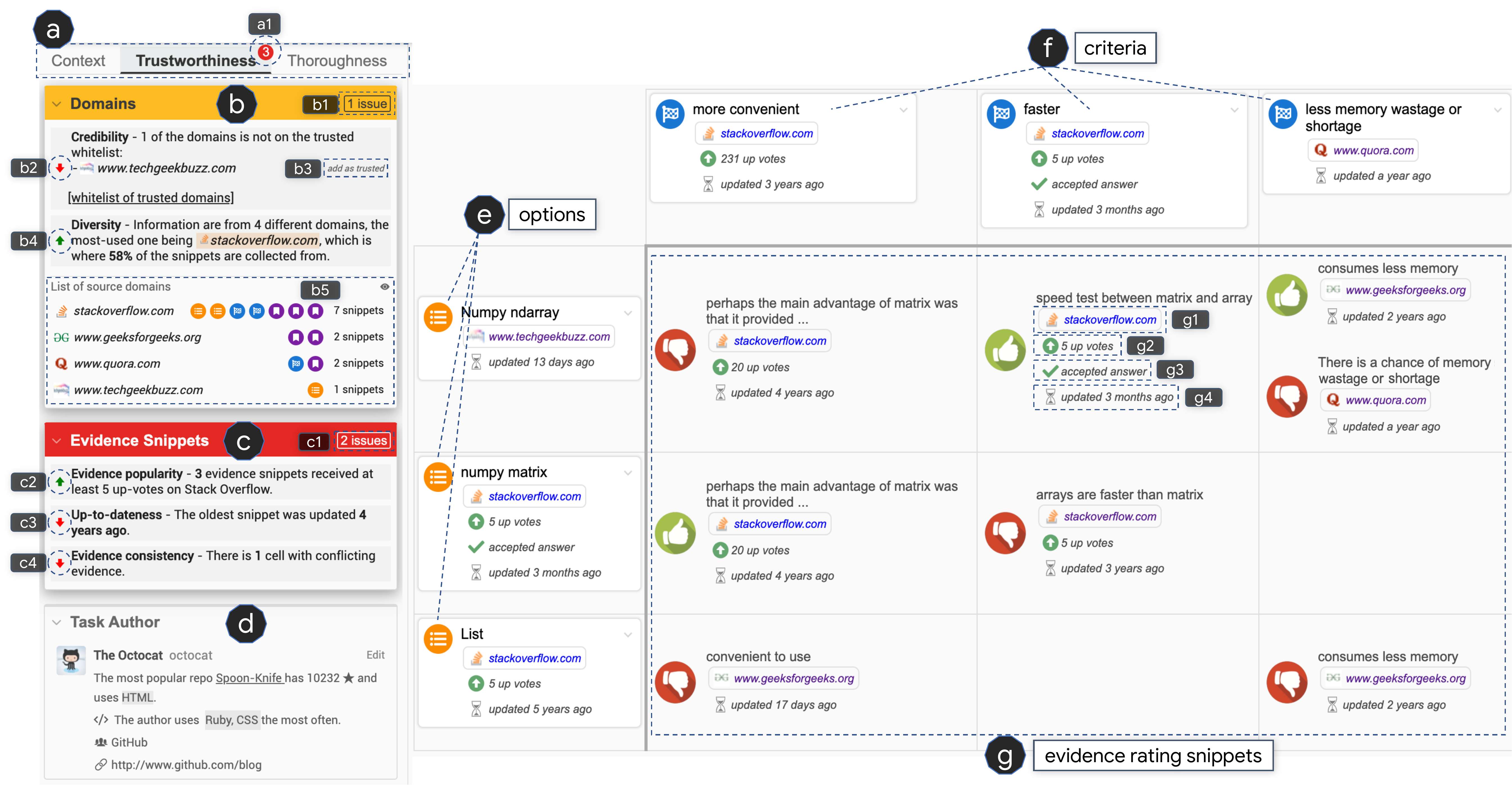}
	\caption{\systemname's user interface. \systemname helps developers evaluate three main facets of appropriateness of reusing a \unakite comparison table with options (e), criteria (f), and evidence (g) through three overview panels: (a) the \textit{Context} panel, the \textit{Trustworthiness} panel, and the \textit{Thoroughness} panel. Each panel contains the \textit{groups} (such as (b), (c), (d)) of appropriateness properties to directly address developers’ information needs. Developers will also be alerted of any potential issues with respect to each facet (e.g., b2, c3, c4).}
	\label{fig:trust-panel}
\end{figure}

There have been various commercial and research information gathering and sensemaking systems that help people with creating reusable knowledge by helping with capturing \cite{bharat_searchpad_2000,alkadhi_rationale_2017,li_sugilite_2017}, organizing \cite{hahn_bento_2018,hahn_knowledge_2016,chang_mesh_2020}, disseminating \cite{chen_codeon:_2017,ponzanelli_seahawk:_2013,li_appinite_2018}, and understanding  \cite{morris_searchtogether:_2007,paul_cosense:_2009,morris_searchbar:_2008,chang_mesh_2020,jin_reviewcollage_2014} information. One that is relevant to the context of programming, our \unakite system \cite{liu_unakite:_2019}, enables developers to collect and organize information online into comparison tables with options, criteria, and evidence to help with making decisions (see Figure \ref{fig:trust-panel}-e,f,g). Systems like these often support keeping track of information for sharing with others later \cite{liu_unakite:_2019,hahn_bento_2018,morris_searchtogether:_2007,paul_cosense:_2009,brandt_example-centric_2010}. For example, \unakite might present a comparison table authored by an initial developer (who we call the \textit{author}) to help subsequent developers (who we call the \textit{consumers} or \textit{readers}) pick an API to represent matrices in Python (as in Figure \ref{fig:trust-panel} and Figure \ref{fig:original-unakite-figure}), or to choose the best \javascript framework to build a website. \unakite is designed to help consumers reuse the decisions and trade-offs identified by the author \cite{liu_unakite:_2019,hsieh_exploratory_2018,gizas_comparative_2012,lawrence_comparing_2017,rutar_comparison_2004} instead of spending the time to discover them from scratch.

However, a major challenge to such a knowledge artifact actually being suitable for reuse is that the consumers do not know if it is \textit{appropriate} to use it or not \cite{markus_toward_2001,watson_multi-theoretical_2006}. Prior research suggested that when checking if a piece of online information can be reused or not, people primarily focus on verifying its \textit{correctness}, and often use \textit{credibility} as a surrogate for correctness because it is easier to check and is highly correlated with correctness \cite{hoorn_web_2010}. For example, signals that can be leveraged to judge credibility include whether the information came from credible sources, whether the way it was presented looked credible, and what the author's qualifications and credentials were \cite{metzger_making_2007,wang_overview_2005,seckler_trust_2015,fogg_elements_1999}. In addition, the correctness of information can not always be measured objectively, but rather often depends on the situation \cite{flanagin_perceptions_2000}; for example, a statement that a sorting algorithm is ``fast'' may depend on the size of the data it is applied to. Furthermore, the knowledge artifacts shown by previous systems are usually a collection and synthesis of different individual pieces of information from different sources. They often capture the author's \textit{opinion} about whether a decision should be made in one way or another, and there is not likely to be a single correct answer but multiple valid options with trade-offs \cite{liu_unakite:_2019}. Unlike general web pages and their content, such knowledge artifacts require many more types of judgements in addition to credibility for someone to decide whether it is appropriate to reuse them or not, including whether the goal and context of the author matches that of the consumer's \cite{markus_toward_2001,hoorn_web_2010}, how thorough was the author's research \cite{paul_cosense:_2009,dourish_awareness_1992}, etc.

Another challenge identified in sensemaking research is that, in reality, consumers often opt to start from scratch rather than reusing previous users' work because of the high costs associated with 1) systematically identifying all of the potential aspects of the work to verify, and 2) obtaining access to properties that could help with the verification \cite{fisher_distributed_2012,lim_assessing_2009,lim_toolkit_2010,metzger_making_2007}. For example, when checking the thoroughness of an author's research, the list of search queries used, the web pages visited, the pages that the author spent the most time reading, and the potential alternatives that were overlooked can all be valid properties to help with the assessment, but are currently not kept track of (even by systems such as \unakite) and hence are not available to the consumer.


In this work, we explore these challenges in the context of reusing the comparison tables created using the \unakite system, where the consumer developer needs to evaluate the \textit{appropriateness} of reusing the table authored by the initial developer.
We perform our research through a \textit{user-centered design} approach. From the vast body of prior work discussing frameworks and measurements for issues of trust and reuse, we extracted properties of importance to developers. We then conducted formative needs-finding interviews with developers about their information needs when evaluating appropriateness. We then synthesized all this information together, resulting in the three key facets of the author's \textit{context}, and the \textit{trustworthiness} and \textit{thoroughness} of the resulting knowledge artifacts, each with a collection of the consumers' specific information needs, which are summarized in a framework in Table \ref{tab:framework-table}. We then devised various key signals and properties that can be used to address those needs as well as mechanisms to automatically identify, compute or keep track of them as an author collects information, which are summarized in the last column of Table \ref{tab:framework-table}. Then, we iteratively designed a hierarchical presentation of the information that lets consumers view and explore those signals and properties interactively, by augmenting the original \unakite tables. These were implemented in a prototype system called \systemname\footnote{\systemname is named after a series of layers of rock that shows the history of a geographical location. It stands for ``\systemnameInFull''.}, which consists of a browser plugin for Google Chrome and a web application (see Figure \ref{fig:trust-panel}). Finally, we conducted a user study to evaluate \systemname's effectiveness.

The primary contributions described in this paper include:
\begin{itemize}
	\item a formative \textbf{study} showing developers' needs for support with reusing previously-generated knowledge,
	\item a synthesized \textbf{framework} (Table \ref{tab:framework-table}) for augmenting judgements of appropriate reuse including three major facets: context, trustworthiness, and thoroughness,
	\item a prototype \textbf{system} called \systemname that automatically records, computes, and visualizes many of the appropriateness signals described in the framework,
	\item an \textbf{evaluation} of the prototype system that offers insights into its usability, usefulness, and effectiveness.
\end{itemize}

\newlength{\rowpaddingbottom}
\setlength{\rowpaddingbottom}{1.2mm}
\begin{table}[h]
\centering
\resizebox{1\textwidth}{!}{%
\begin{tabular}{c|p{25mm}p{51mm}p{51mm}p{51mm}}
    \toprule
    \textbf{Facet} & 
    \textbf{Information Need} &
    \textbf{Selected References\newline in Prior Research} & 
    \textbf{Sample Quotes in\newline Formative Study} & 
    \textbf{Selected Supporting Features\newline in \systemname} \\
    \midrule
    
    \multirow{3}{*}{\vspace{-45mm}\rotatebox[origin=c]{90}{\textbf{Context}}} & 
    Goals of the\newline original decision &
    \dottedPriorWork{Search queries are useful for encoding task goals \& contexts in various settings like asynchronous collaborations \cite{bharat_searchpad_2000,morris_searchtogether:_2007,paul_cosense:_2009,paul_sensemaking_2011,sharma_artifact_2009,yue_investigation_2012}.}\vspace{\rowpaddingbottom} & 
    \dottedExp{``This looks like it's trying to pick a speech recognition API, but what I want is actually text to speech.''}\vspace{\rowpaddingbottom} &
    \dottedPriorWork{Keeping track of the author's search queries to reflect his or her task goal.}
    \\
    
    & 
    Explanation or\newline 
    contextualization of\newline information &
    \dottedPriorWork{Recontextualization of information helps with understanding \cite{markus_toward_2001,liu_unakite:_2019}.}\par \dottedPriorWork{Clarity and informativeness of website content improves understanding \cite{fogg_persuasive_2002,thielsch_facets_2019}.}\vspace{\rowpaddingbottom} &
    \dottedExp{``What does this `very efficient' mean, is it `memory' or `time' efficient?''}\par \dottedExp{``Is it [a sorting algorithm] `fast' only when there're a few hundred data points or also when there are millions of data points?''}\vspace{\rowpaddingbottom} & 
    \dottedPriorWork{Keeping track of the surroundings along with the information snippets and presenting them as contextual explanations.}
    \\
    
    & 
    Situational\newline awareness &
    \dottedPriorWork{Awareness of common ground facilitates sensemaking handoff \cite{sharma_artifact_2009,sharma_sensemaking_2008,clark_grounding_1991}.}\par \dottedPriorWork{Users need awareness of each others' actions in order to perform their tasks better \cite{myers_flexi-modal_2002,morris_searchtogether:_2007,paul_cosense:_2009,amershi_cosearch:_2008}.} &
    \dottedExp{``I want to solve it with pure JavaScript, but it seems that most of the answers here are actually written using jQuery?''}\par \dottedExp{``I'm using Python 2.7 at the moment, which is fairly old, does this example also use this version?''} & 
    \dottedPriorWork{Detecting information about languages, frameworks, and their versions mentioned in information snippets with a predefined yet easily extensible list of detectors.}
    \\ 
    \midrule
    
    \multirow{5}{*}{\vspace{-75mm}\rotatebox[origin=c]{90}{\textbf{Trustworthiness}}} & 
    Source credibility\newline and diversity &
     \dottedPriorWork{Source credibility affects trustworthiness of information \cite{denning_wikipedia_2005,eysenbach_how_2002,fogg_persuasive_2002,metzger_making_2007,thielsch_facets_2019}.}\par \dottedPriorWork{Sources similar to what a consumer usually uses are more likely to be deemed credible \cite{seckler_trust_2015,metzger_social_2010}.}\vspace{\rowpaddingbottom} &
    \dottedExp{``If it's from Stack Overflow, I'm usually fine with it. But if it's from some random blog posts written by some random guy, I would think twice.''}\par \dottedExp{``I wonder if all of these just came from the official documentation or there're also other developer forums.''}\vspace{\rowpaddingbottom} &
    \dottedPriorWork{Visualizing the distribution of information snippets across different domains (websites).} \par \dottedPriorWork{Alerting consumers of potential untrusted domains.}
   \\
    
    & 
    Information \newline up-to-dateness &
    \dottedPriorWork{Information currency affects its perceived credibility \cite{alexander_web_1999,brandt_evaluating_1996,metzger_making_2007}.}\vspace{\rowpaddingbottom} &
    \dottedExp{``Is this speed comparison [between React, Angular, and Vue] up-to-date now that Angular 9 was just released?''}\vspace{\rowpaddingbottom} &
    \dottedPriorWork{Extracting and surfacing the last updated time of information snippets.}
    \\
    
    & 
    Information\newline popularity & 
    \dottedPriorWork{People apply \textit{the endorsement heuristic} to evaluate credibility \cite{metzger_making_2007}.}\par \dottedPriorWork{People seek social proof when evaluating credibility \cite{seckler_trust_2015}.}\vspace{\rowpaddingbottom} &
    \dottedExp{``If there're a lot of other devs [who] also think this is a good idea, then I'm much more comfortable to use it.''}\vspace{\rowpaddingbottom} & 
    \dottedPriorWork{Extracting and surfacing signals showing information popularity, such as the up-vote count of an answer on \stackoverflow.}
    \\ 
    
    & 
    Information\newline consistency & 
    \dottedPriorWork{People apply \textit{the consistency heuristic} to evaluate credibility \cite{metzger_making_2007}.}\par \dottedPriorWork{People seek more than one source to verify information \cite{meola_chucking_2004}.}\vspace{\rowpaddingbottom} &
    \dottedExp{``It claims PyTorch is much easier to learn than Tensorflow, but I wonder if there're people suggesting otherwise.''}\vspace{\rowpaddingbottom} &
    \dottedPriorWork{Alerting consumers if there are conflicting  (both positive and negative) ratings in any of the table cells.}
    \\ 
    
    & 
    Author credibility & 
    \dottedPriorWork{The author's level of expertise affects information trustworthiness \cite{denning_wikipedia_2005,seckler_trust_2015,kittur_can_2008}.}\par \dottedPriorWork{Disclosing patterns of past performance helps people evaluate trustworthiness \cite{kittur_can_2008,suh_lifting_2008,shneiderman_designing_2000}.} &
    \dottedExp{``Does the table author know what he's doing?''} \par \dottedExp{``Is the author saying all the nice things about Caffe because he has lots of experience with it or because he's biased?''} & 
    \dottedPriorWork{Surfacing credibility and bias signals from the table author's Github profile, such as their primary programming language, number of stars on their repositories, and affiliation.}
    \\
    \midrule

    \multirow{3}{*}{\vspace{-40mm}\rotatebox[origin=c]{90}{\textbf{Thoroughness}}} & 
    Research process and effort &
    \dottedPriorWork{External representations handed off should indicate prior investigative process and insights \cite{paul_sensemaking_2011,paul_cosense:_2009,zhao_supporting_2018}, how much work had been done, and how mature the representation was \cite{sharma_artifact_2009,sharma_sensemaking_2008}.}\vspace{\rowpaddingbottom} &
    \dottedExp{``How much effort was put into making this decision?''}\par \dottedExp{``What did the author focus on?''}\vspace{\rowpaddingbottom}  &
    \dottedPriorWork{Keeping track of and visualizing the author's activities on an interactive timeline view, including search queries, pages visited, duration of stay on the pages, information snippets collected, etc.}
    \\
    
    & 
    Alternatives or\newline competitors &
    \dottedPriorWork{Knowledge and sensemaking results should indicate their coverage and scope \cite{denning_wikipedia_2005,metzger_making_2007}.}\vspace{\rowpaddingbottom} &
    \dottedExp{``I heard anecdotally that Svelte gives you much better performance than all these big (\javascript) frameworks [React, Angular, and Vue]. I should take a look at that before I decide.''}\vspace{\rowpaddingbottom} & 
    \dottedPriorWork{Finding and surfacing commonly searched-for alternatives mentioned in Google autocomplete suggestions.}
    \\
    
    & 
    Usable artifacts & 
    \dottedPriorWork{Developers need help finding and reusing code examples \cite{brandt_example-centric_2010,oney_codelets:_2012,ponzanelli_seahawk:_2013}.} &
    \dottedExp{``Which option was chosen in the end?''}\par \dottedExp{``[Are there] any code snippets that I can immediately plug into mine and test?''} & 
    \dottedPriorWork{Extracting and surfacing code examples from information snippets.}
    \\ 
    \bottomrule
\end{tabular}%
}
\caption{A framework summarizing the three major facets (column 1) when evaluating the appropriateness to reuse knowledge, including people's specific information needs (column 2), selected evidence from prior work (column 3), sample quotes from our formative study interviews (column 4), and features we devised to support the information needs in the subsequent \systemname system (column 5).}
\label{tab:framework-table}
\vspace{-16.9mm} 
\end{table}

%
%
%
%
%
%
\section{Related Work}
\label{sec:related-work}
\subsection{Information and Knowledge Reuse}
As formulated by Davenport et al. in 1996 \cite{davenport_improving_1996} and Markus in 2001\cite{markus_toward_2001}, knowledge processes are often categorized by whether they involve \textit{knowledge creation} (e.g., research and development of new products and services, or writing books or articles) or \textit{knowledge reuse} (e.g., reapplying existing components and best practices to solve common problems). While there is much research into the significance and difficulties of knowledge creation and innovation \cite{davenport_improving_1996,kittur_standing_2014,nonaka_theory_1996,krogh_enabling_2000,hahn_knowledge_2016,hahn_knowledge-based_1999}, the effective reuse of knowledge has been shown to be a more frequent strategy and concern to individuals and organizations \cite{markus_toward_2001,davenport_improving_1996,dixon_common_2000,odell_if_1998,osterloh_motivation_2000,zhang_wikum:_2017,zhang_making_2018}.

Many systems have been developed to support the multiple stages of information and knowledge reuse as mapped out by Markus \cite{markus_toward_2001}: \textit{capturing and documenting knowledge}, \textit{packaging and distributing knowledge}, and \textit{reusing knowledge}. Among them, some systems support capturing, organizing, and keeping track of information in the first place (e.g., \cite{liu_unakite:_2019,bharat_searchpad_2000,hahn_bento_2018,vermette_cheatsheet:_2015,li_kite_2018,li_pumice_2019}), some aim to deliver and surface existing knowledge directly to a user without the need of complex matching and frequent context switches (e.g., \cite{ponzanelli_seahawk:_2013,brandt_example-centric_2010,chang_searchlens_2019}), and others facilitate the digesting and understanding of knowledge (e.g., \cite{liu_unakite:_2019,suh_lifting_2008,lim_toolkit_2010}). However, having a literal understanding of a knowledge artifact does not by itself imply reuse --- a major barrier to that knowledge actually being useful is the consumer does not know whether it is \textit{appropriate} to use it or not \cite{markus_toward_2001,watson_multi-theoretical_2006}. 

Prior research provides insights into various properties that people look for in order to evaluate the appropriateness for reuse, such as source credibility \cite{denning_wikipedia_2005,eysenbach_how_2002,fogg_persuasive_2002,metzger_making_2007,seckler_trust_2015,thielsch_facets_2019}, information currency (or up-to-dateness) \cite{alexander_web_1999,brandt_evaluating_1996,metzger_making_2007}, information popularity \cite{metzger_making_2007,seckler_trust_2015}, goals and purposes (what the author wanted to achieve) \cite{paul_sensemaking_2011,sharma_artifact_2009}, etc. 
However, much research such as the above focuses on specific issues about the general credibility of web content, while knowledge artifacts previously collected and synthesized by an author require many more types of judgements beyond credibility in order for a consumer to decide its appropriateness for reuse. To the best of our knowledge, there remains no systematic models or frameworks for understanding the factors that affect the judgements of the reuse of previously created knowledge artifacts. Such a framework could be helpful for driving research studying and augmenting reuse across a variety of domains and forms. Here we take a step towards such a framework, starting with knowledge artifacts in the form of comparison tables, which are widely used, and in the domain of programming, where knowledge reuse happens frequently \cite{sojer_code_2010,sojer_code_2010,haefliger_code_2007,hesse_documented_2016,brandt_example-centric_2010,hsieh_exploratory_2018,liu_unakite:_2019,ponzanelli_seahawk:_2013,ko_six_2004,latoza_hard--answer_2010}. In the following sections, we discuss three of the most relevant threads of research as they relate to judgements of knowledge reuse.

\subsection{Evaluating Online Information Credibility}
\subsubsection{Models and Heuristics for Evaluating Online Information Credibility}
One of the most researched facets of knowledge reuse is evaluating online information credibility \cite{metzger_making_2007,wang_overview_2005,seckler_trust_2015,fogg_elements_1999} (or ``trustworthiness'' \cite{thielsch_facets_2019}), which focuses on facets of authenticity, reliability, and trustworthiness of a given piece of content online, ranging from e-commerce transactions to online discussions and collaborations \cite{thaw_study_2009,kittur_can_2008,suh_lifting_2008}. Prior work has employed bottom-up approaches like surveys and contextual inquiries and reported various factors that influence credibility assessment, including but not limited to: domain name and URL, presence of date stamp showing information is current, author identification and indication of his or her expertise, citations to scientific data or references, and user ratings and reviews
\cite{wang_overview_2005,meola_chucking_2004,shneiderman_designing_2000,brandt_evaluating_1996,fritch_evaluating_2001,thielsch_facets_2019,eysenbach_how_2002,metzger_making_2007,fogg_persuasive_2002,metzger_social_2010,alexander_web_1999}. 

In addition, models and heuristics for credibility assessment have also been proposed, for example, the \textit{checklist model}, which guides users through a checklist of critical factors during assessment \cite{metzger_making_2007}, and the \textit{contextual model}, which emphasizes the use of external information to establish credibility \cite{meola_chucking_2004}, such as promoting peer-reviewed resources and seeking corroborating or conflicting evidence. A summary by Metzger et al. \cite{metzger_social_2010} suggests that users routinely invoke \textit{cognitive heuristics} to evaluate the credibility of information and sources online, such as the \textit{reputation heuristic} (checking if the source of the information has good reputation and credentials), and the \textit{expectancy violation heuristic} (checking if a website or its content conforms to their original expectations).

However, in reality, it has repeatedly been shown that people are often underprepared and have trouble determining how to evaluate the credibility of online information \cite{amsbary_factors_2003,meola_chucking_2004,metzger_credibility_2003,scholz-crane_evaluating_1998}, which is often deemed to be too much work \cite{meola_chucking_2004,sharma_sensemaking_2008}, having a high possibility of missing important details \cite{metzger_making_2007,metzger_social_2010}, and eventually leading to abandonment, mistrust or misuse \cite{lim_assessing_2009,lim_toolkit_2010,metzger_making_2007} of the information. This reflects a significant gap between research and reality: while prior work provides insights into the various factors affecting online information credibility and ways people reason about them, people need tool support that systematically helps with credibility assessment and information reuse. We address this gap by providing a prototype system that (1) automatically extracts appropriateness signals (including those related to credibility) from the original knowledge content when possible; and (2) processes and presents them to the consumer of the knowledge in a hierarchical visualization that directly addresses their information needs during the evaluation of the appropriateness to reuse.

\subsubsection{Interventions and Support for Evaluating Collaboratively-built Knowledge Content}
Collaborative knowledge building, exemplified by the Wikipedia project \cite{noauthor_wikipedia_nodate} and \stackoverflow \cite{noauthor_stack_nodate}, has become highly popular in many domains, and its mutable nature that virtually \textit{anyone can edit anything} has invited considerable research into helping users evaluate the trustworthiness of its content. For example, the revision histories \cite{suh_lifting_2008, viegas_studying_2004,zeng_computing_2006,zeng_mining_2006}, review processes \cite{viegas_hidden_2007}, and the external references \cite{fogg_elements_1999,fourney_enhancing_2013} of an article can be modeled and visualized to help improve transparency and the evaluation of its trustworthiness. In addition, an author's past performance, such as their editing history on Wikipedia or previously answered questions on \stackoverflow, can be mined \cite{adler_content-driven_2007,shneiderman_designing_2000} and surfaced \cite{suh_lifting_2008} to help knowledge consumers determine the author's reputation, expertise, and other accountability metrics. Encouragingly, Kittur et al. \cite{kittur_can_2008} showed that surfacing trust-relevant information from Wikipedia articles had a dramatic impact on users' perceived trustworthiness of those articles, holding constant the content itself.

However, despite the overwhelming importance and increasing research effort, being considered trustworthy is often not the sufficient condition for reuse, \revision{nor is trustworthiness always the first facet that users evaluate} --- research has shown that people often have trouble understanding a piece of information when it is taken out of its original context \cite{liu_unakite:_2019,markus_toward_2001} and figuring out if it is indeed relevant to their own situation \cite{saracevic_relevance_nodate,borlund_concept_2003,sharma_sensemaking_2008} before they start to think about trustworthiness and credibility. In addition, they also wonder about how much effort has been put into creating a piece of knowledge and does it cover everything that they are interested in \cite{sharma_artifact_2009,sharma_sensemaking_2008,paul_cosense:_2009,zhao_supporting_2018,markus_toward_2001} before they can give a final verdict on reusing it or not. Therefore, we draw from and build upon these prior works, where we iterated to identify, extract, and surface not only the important elements of trustworthiness but also context and thoroughness to help consumers make a more comprehensive assessment of the appropriateness of reusing knowledge, exemplified by decisions and their rationale in programming.

\subsection{Sensemaking Handoff}
Much research has explored the activity of \textit{sensemaking handoff}, during which one individual must continue the sensemaking work where another has left off. It frequently happens in asynchronous collaborations \cite{paul_cosense:_2009,paul_sensemaking_2011,zhao_supporting_2018,fisher_distributed_2012}, shift changes \cite{patterson_shift_2001}, etc., during which the current sensemaker (consumer) needs to make sense of and evaluate the appropriateness of reusing the results generated by a previous sensemaker (author) \cite{sharma_sensemaking_2008,markus_toward_2001}. Various metadata and properties parallel to the main artifacts of sensemaking have been proposed that would help the people with this process, such as the awareness of the previous sensemaking process \cite{paul_cosense:_2009,dourish_awareness_1992} (e.g., search queries and visited web pages), the level of expertise of the author \cite{markus_toward_2001,sharma_artifact_2009}, and the context of the original sensemaking problem \cite{markus_toward_2001}.

However, it is both time and effort intensive for an author to keep track of their rationale and processes with little immediate payoff, which is also often for the benefit of others rather than themselves \cite{liu_unakite:_2019}. Even in situations where authors have the explicit wish to help, they are often uncertain of what metadata and properties to provide and how those can be instantiated using concrete signals that would be valuable to the consumers in evaluating the reusability of their sensemaking results \cite{sharma_sensemaking_2008}. We address these barriers in the context of reusing decisions in programming by iteratively developing a framework that summarizes the major facets that consumers care about during the evaluation of appropriateness to reuse along with the corresponding detailed information signals, and a set of technical approaches that can automatically extract, compute, and visualize them when possible. We integrated these into our \unakite system \cite{liu_unakite:_2019} that helps authors organize and record their decisions for reuse, saving them the burden of coming up with the appropriate signals to keep track of as well as potential extra effort needed to accurately obtain them.

\begin{figure}[t]
\centering
	\includegraphics[width=\linewidth]{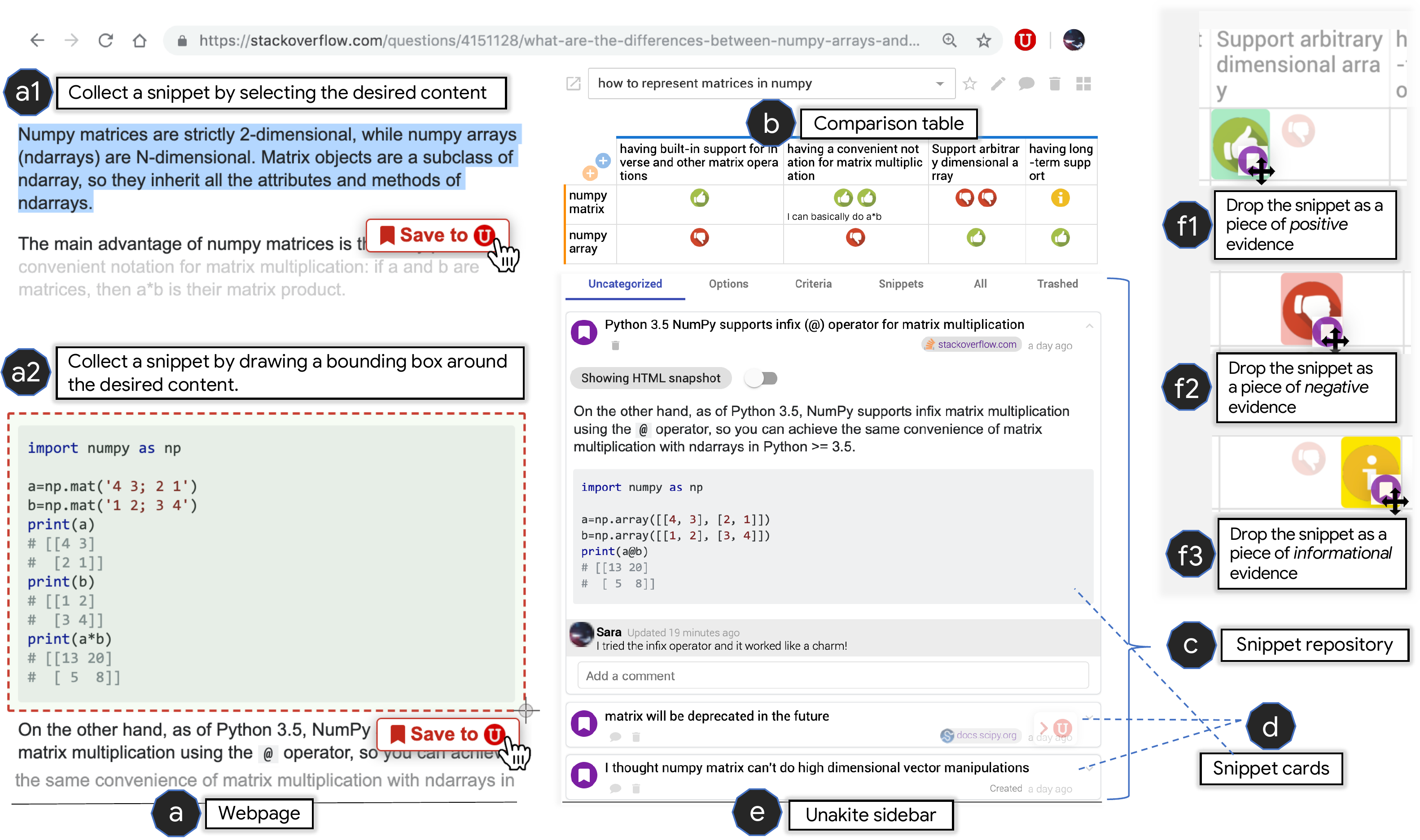}
	\caption{\unakite's user interfaces. With \unakite, a developer collects a snippet by selecting the desired content (a1) or by drawing a bounding box around the desired content (while holding the \texttt{Option} / \texttt{Alt} key) (a2) and clicking the ``Save to U'' button. The collected snippet will show up under the ``Uncategorized'' tab in the snippet repository (c) as a snippet card (d) inside the \unakite sidebar (e), which shows the current task at the top (``how to represent matrices in numpy''). The developer can drag the snippet and drop it in one of the cells in the comparison table near the top (b), and mark whether it is positive (green thumbs-up) or negative (red thumbs-down) or just informational (yellow ``i''). (f1-f3) show the details of the three parts of each cell in the table where the snippet can be dropped. For full details, see \cite{liu_unakite:_2019}.}
	\label{fig:original-unakite-figure}
\end{figure}

\subsection{Knowledge Reuse in Programming}
The practice of knowledge reuse has been particularly relevant in the software industry \cite{haefliger_code_2007}. Code reuse, in particular, has become a hugely successful paradigm in the development of new software products and services in both the commercial and open source sector. Developers frequently use well-maintained functional code modules from code-sharing platforms such as GitHub \cite{noauthor_build_nodate} and npm \cite{noauthor_npm_nodate}, enjoying the benefits of significantly reduced workload, improved productivity, enhanced software performance, stability and security, and more time for innovation \cite{haefliger_code_2007,sojer_code_2010,markus_toward_2001,mockus_large-scale_2007,frakes_software_1986,frakes_software_1996,myers_making_2017,jin_providing_2020}.

Despite the fact that software code is the most obvious target for reuse \cite{sojer_code_2010,mockus_large-scale_2007,haefliger_code_2007}, knowledge reuse in programming may go well beyond code, as stated by Barns and Bollinger \cite{barns_making_1991}: ``The defining characteristic of good reuse is not the reuse of software \textit{per se}, but the reuse of human problem-solving.'' Indeed, developers on community Q\&A websites like \stackoverflow \cite{noauthor_stack_nodate} share not only code examples \cite{brandt_example-centric_2010,ponzanelli_seahawk:_2013} but also decision making strategies, design rationale such as alternative options, criteria or constraints that should be met, and the resulting trade-offs \cite{liu_unakite:_2019,hsieh_exploratory_2018}. Furthermore, questions about design rationale are widely cited by developers as some of the hardest to answer \cite{latoza_hard--answer_2010,latoza_maintaining_2006,sillito_questions_2006}. Tools like \unakite \cite{liu_unakite:_2019} can greatly reduce the costs to keep track of and later understand such rationale knowledge, with the hope that such knowledge can ultimately be better reused rather than be obtained from scratch requiring duplicated research effort \cite{haefliger_code_2007,liu_supporting_2018}. In the current work, we further advance this research thread by developing features and affordances enabling developers to evaluate the context, trustworthiness, and thoroughness of previously-made decisions, which is arguably one of the missing links between understanding and reuse.

%
%
%
%
%
%
\section{Background and Formative Investigations}
\label{sec:formative}

In this work, we explore augmenting knowledge reuse judgements in the context of programming, specifically in using the \unakite \cite{liu_unakite:_2019,liu_unakite_2018} system. We first explain the design and usage of \unakite to provide a background for our research, and then describe a formative study investigating developers' issues and information needs for knowledge reuse when using \unakite.

\subsection{The \unakite System}
As mentioned, \unakite addresses both the need of initial developers to synthesize online information and recognize the trade-offs in programming decisions and the need of subsequent developers to be able to understand the rationale behind those decisions. \unakite, as a Chrome extension, enables the initial developers to easily collect any content from any web page as \textit{snippets} (pieces of information, Figure \ref{fig:original-unakite-figure}-d) into the snippet repository (Figure \ref{fig:original-unakite-figure}-c) by either selecting or dragging out a bounding box around the desired content using the cursor (Figure \ref{fig:original-unakite-figure}-a1,2). To help with organization, developers can use drag-and-drop to move their collected snippets into a comparison table (Figure \ref{fig:original-unakite-figure}-b) with \textit{options} (as row headers), \textit{criteria} (as column headers), and \textit{evidence} (``thumbs-up'' or positive, ``thumbs-down'' or negative, and ``informational'' (``i'') ratings that spread across the rest of the table cells) that illustrates the trade-offs among various solutions. All the interaction techniques involved are designed to be natural and lightweight without taxing users with much cognitive load \cite{kittur_standing_2014,kittur_costs_2013} while they are searching and exploring for potential solutions to their programming problems.

The resulting organizational structure is automatically saved by the system and can be accessed through a web application (with the snippet repository on the left and the comparison table on the right) with a unique URL, which can be used as stand-alone documentation of the design rationale or be integrated with code through comments. As opposed to having to speculate about the correctness and legitimacy of a decision \cite{ko_information_2007}, subsequent developers who have access to these comparison tables will be able to understand the context of the decision space: what options and alternatives were considered, what criteria or constraints should be met, what the resulting trade-offs were, and what was deemed to be the most important and why.

Although \unakite has been shown through lab studies \cite{liu_unakite:_2019} to help the initial developer in making a programming decision, it displays few of the signals suggested by the research discussed above on trust and sensemaking handoff that could help consumers of the table decide whether it is appropriate for them to reuse it. For example, the initial table creator may or may not have been thorough in their research; may or may not have the same context and environment; or may or may not care about the same goals as the consumer. Although we use \unakite as a specific context, there are many similar examples of developers creating comparison tables in code documentation, blogs, and \stackoverflow \cite{noauthor_put_2009,noauthor_which_2009}, which are typically even sparser in terms of signals for reuse appropriateness, with no supporting interactivity or drill-downs possible.

\subsection{Formative Interviews}

To characterize the prevalence and types of issues developers have with knowledge reuse, specifically with reuse of programming decisions, we conducted semi-structured interviews with 15 developers (5 female, 10 male). Participants were recruited through mailing lists, social media postings, and word-of-mouth. To capture a variety of processes, we chose 8 professional developers, 3 doctoral students, and 4 master students. While we do not claim that this sample is representative of all developers, the interviews informed and motivated the development of the subsequent framework (Table \ref{tab:framework-table}) and the design of the \systemname system. 

We began by asking participants about their experiences in reusing someone else's decisions when programming and how frequently would that situation occur in their work. We then explored how they manage these situations and their information needs, in particular, what questions do they have when evaluating the appropriateness to reuse and how answers to those questions may affect their final verdicts on reusability. In addition to eliciting facts on their past experiences, we also presented them with a set of decision tables in the running \unakite application (which were directly adapted from real tables online, e.g., \cite{score_why_2017}) as well as the corresponding background situational context, and asked them to judge if they could reuse these tables in those given situations. We asked them to speak about any questions they had and perform any inquiry they wanted to answer those questions (e.g., checking the sources, searching for evidence online, etc.). Finally, we wrapped up with questions probing their experience with explaining their design rationale to others, and whether and how do they convince others that their decisions are appropriate to be reused.

Interviews were conducted either in person or remotely by the first author and lasted 30 minutes. They were audio-recorded and then transcribed. In addition, screenshots of participants' computers were taken for later analysis when applicable. Then, the first author went through the transcriptions and coded them via an open coding approach \cite{charmaz_constructing_2006}, which included multiple iterations of discussions with the research team. Our key findings are presented below.

\subsection{Preliminary Results}

\subsubsection{Decision reuse is frequent in programming.}
All participants were able to recall and describe experiences of evaluating and reusing someone else's decisions. One of the scenarios where reuse frequently happens is during code refactoring, makeovers, and takeovers, where developers are required to re-evaluate decisions made by some other developers or teams for reusability. For example, P8, a professional full-stack developer, said: \textit{``just last month we were taking over another team’s project, and the first thing we did was to re-evaluate if it still makes sense to continue building with Ruby on Rails, or it’s time to do a whole re-write with React or Angular.''} Another frequent reuse scenario is during project startups, where developers actively look for existing decisions on choosing architecture, frameworks, libraries, algorithms, and APIs to reuse, such as \textit{``picking the right cryptographic algorithm to encrypt passwords''} (P12) and \textit{``choosing the best optimization method to train neural networks''} (P6).

\subsubsection{Developers need guidance and tool support when evaluating whether to reuse someone else's decision.}
Although decision reuse happens frequently, participants' strategies to evaluate the appropriateness of reuse were not without troubles. 10 out of 15 said that they usually have some ideas of what types of evidence to look out for, such as the credibility of the sources, code examples, and library version mismatches, etc., but were not confident that those were sufficient. For example, P5 said: \textit{``I feel the obligation to do more validations other than confirming they [performance metrics for different deep learning frameworks] are from official docs, but I’m not sure what else to look at or where I can find extra information.''} In addition, participants reported often having to manually look for evidence of reusability (7/15), such as following the URLs previous developers left in code comments to the original web pages to validate information correctness, and pinging and asking the original author about what alternatives were considered back then. However, sometimes the sources of evidence were not possible to find since none of the sensemaking processes were consciously kept track of during the original author's decision making process other than the result.

%
%
%
%
%
%
%
%
\section{Framework}
\label{sec:framework}
Data from the formative study suggested that developers would benefit from support in evaluating the appropriateness of reusing decisions. For example, there are many indicators that could be beneficial to surface to help users make these judgements, ranging from the expertise of the author to the \revision{quantity} and legitimacy of the sources used. Although there has been little prior work characterizing the most important factors for decision reuse specifically by developers, as listed above there has been significant work discussing frameworks and measurements relevant to evaluating and reusing knowledge, such as online information credibility judgement \cite{metzger_making_2007,metzger_social_2010,meola_chucking_2004,metzger_credibility_2003}, asynchronous collaboration \cite{paul_sensemaking_2011,morris_searchtogether:_2007}, and sensemaking handoff \cite{sharma_sensemaking_2008,sharma_role_2011,sharma_artifact_2009,fisher_distributed_2012}. From these research papers, we extracted properties and signals that would be important and relevant to decision reuse for developers. 

By coding and synthesizing the aforementioned prior work as well as the formative study results through affinity diagramming, we identified three major clusters, that we call \textit{facets}, when evaluating the appropriateness for reuse in programming: the original author's decision making \textit{context}, and the \textit{trustworthiness} and \textit{thoroughness} of the resulting decision. We used these as a guide in developing an integrated framework, shown in Table \ref{tab:framework-table}, consisting of the three identified facets (column 1), specific information needs of developers with regard to each facet (column 2), selected evidence for the importance of these information needs as well as possible solutions to address them from prior work (column 3), and sample quotes from our formative interviews (column 4). These insights together inspired the features for our subsequent \systemname system (column 5). We now discuss the framework in detail, along with the support from the prior work and the formative interviews. The design of \systemname follows in section \ref{sec:design-and-implementation}.

\subsection{Context}
Although in prior work the importance of understanding the trustworthiness of information often outshines everything else when evaluating the appropriateness to reuse \cite{markus_toward_2001,kittur_can_2008}, we were surprised to find out that, at least in the domain of programming decision reuse, developers often ask questions about the \textit{context} of a previously-made decision before they proceed to assess trustworthiness (9/15). Cited reasons include that one needs to know \textit{``how relevant it is to what I am doing''} (P5) first, and if the context of the original decision does not align very well with the problem at hand, one would often stop the evaluation process and move on to look for new solutions. For example, if a developer is working in Java, solutions that only work in \revision{\javascript may not be worth investigating}.

\subsubsection{Goals of the original decision}
When evaluating context, most (12/15) participants asked questions about the goals and purposes of the author of the decision in order to compare those with their own. For example, \textit{``this looks like it's trying to pick a speech recognition API, but what I want is actually text to speech,''} (P14) and \textit{``people say they want to do one thing, but after taking a closer look, they really are doing this other thing, which often makes me a tad frustrated''} (P7). Indeed, prior research suggests that the goals of decisions are often treated as ``self-evident'' given the results, and therefore are often not kept track of by the authors \cite{latoza_hard--answer_2010,latoza_maintaining_2006}. On the other hand, goal mismatch does not always prevent developers from further evaluating a decision; instead, it can become a \textit{``learning opportunity''} for them to \textit{``know more about a new technology or design pattern''} (P11). 

Furthermore, when asked about their experience of making decisions, participants reported that their goals may very well evolve with their exploration process rather than remaining \revision{fixed} from the beginning (7/15). For example, \textit{``I started out trying to choose a framework to build a mobile app for both Android and iOS, but later I stumbled upon this progressive web app thing that totally fulfills all of my requirements, so I ended up trying to learn more about that, and sort of abandoned the mobile app route that I was originally planning to take''} (P3). This motivated us to develop features (e.g., keeping track of all of the search queries used) to capture not only an author's original goal but also the evolving nature of that goal, so that later knowledge consumers could have a better grasp of how the author's goal changed throughout a decision making process.

\subsubsection{Explanation or contextualization of information}
One of the frustrations that participants reported having is that they often have trouble understanding the meaning of some of the criteria and evidence used in online decision tables (8/15). For example, \textit{``what does this `very efficient' mean, is it `memory' or `time' efficient?''} (P10). In some other circumstances, they suspect that evidence may not hold true when external constraints or requirements change: \textit{``is it [a sorting algorithm] `fast' only when there're a few hundred data points or also when there are millions of data points''} (P1). Indeed, prior work suggests that clarity and informativeness of information have a significant impact on how well it is understood \cite{fogg_persuasive_2002,thielsch_facets_2019}, and presenting information along with its original context (recontextualization) is considered a good way to help people understand its meaning and the conditions in which it is correct or accurate \cite{liu_unakite:_2019,markus_toward_2001,flanagin_perceptions_2000}. 

In addition, it was also suggested by participants that it is not always easy to recontextualize information, especially when the context is not available (6/15). \unakite partially addressed this by allowing users to create a snippet out of a large block of information in its original HTML format as well as automatically recording the corresponding source URL for later retracing \cite{liu_unakite:_2019}. In \systemname, we build on that by introducing the concept of a \textit{context snapshot}, which, at capture time, automatically keeps track of the \textit{surroundings} of an information snippet in addition to the snippet content itself and its source URL. When consumers are reviewing a snippet, they will be able to benefit from the possible explanations such as code examples and performance metrics contained in the surroundings that would otherwise be missing from the snippet content.

\subsubsection{Situational awareness} 
An essential part of context is the situation in which the information will be reused. In programming, this corresponds to the languages, libraries, and platforms being used, which are often referred to as \textit{dependencies}, and participants reported checking if a given decision shares the same language or library usage as to what they have to work with (8/15). For example, P7 asked \textit{``I want to solve it with pure \javascript, but it seems that most of the answers here are actually written using jQuery.''} Furthermore, version mismatch has been a frequent issue for reuse in programming. With the continuous rise of the open source software development model \cite{haefliger_code_2007} and the increasing number of frameworks, libraries, languages, and patterns \cite{noauthor_front-end_2019,noauthor_getting_nodate,noauthor_programming_nodate}, version and dependency mismatches and errors can cause troubles from missing features to breaking dependent downstream applications \cite{noauthor_exports_2020}. Indeed, participants reported checking for versions before they commit to adopting a certain solution (6/15). For example, \textit{``I'm using Python 2.7 at the moment, which is fairly old; does this example also use this version, or is it using Python 3.5?''} These inspired us to try to automatically detect the language, library, platform, and version information whenever possible when an author collects information online, and surface this to the consumer to directly address their information needs.

\subsection{Trustworthiness}

As mentioned, information trustworthiness or credibility is often used as a surrogate for verifying information correctness \cite{hoorn_web_2010}, and is one of the most reported and researched facets during the evaluation of the appropriateness to reuse knowledge across many domains \cite{markus_toward_2001,metzger_making_2007}. Our interview data shows that it plays a crucial role in the domain of reusing decisions in programming as well.

\subsubsection{Source credibility and diversity}
As suggested by prior work, source credibility has a significant impact on the trustworthiness of information \cite{denning_wikipedia_2005,eysenbach_how_2002,fogg_persuasive_2002,metzger_making_2007,thielsch_facets_2019}. Not surprisingly, all participants in our study reported this same belief --- they are more inclined towards trusting information from sources that are official (e.g., API documentation websites) or with a very good reputation within the community (e.g., \stackoverflow), and are more likely to reject information from sources that they have little experience with, echoing the \textit{reputation heuristic} and the \textit{expectancy violation heuristic} \cite{metzger_social_2010,seckler_trust_2015} that people generally use to assess trustworthiness. For example, P12 said: \textit{``if it's from \stackoverflow, I'm usually fine with it. But if it's from some random blog posts written by some random guy, I would probably think twice.''}

It is worth noting that in addition to credibility, source diversity also plays a role in trustworthiness, according to 7 of the 15 participants. They thought that the more diverse the sources used are, the more likely that the evidence in the table has been \textit{``peer reviewed''} or \textit{``confirmed by a bunch of other devs''}, and \textit{``seeing essentially the same thing independently said on a couple of different sites and forums''} \revision{gives} them \textit{``peace of mind''}. We believe that source diversity also works in concert with information popularity and consistency, which we will discuss in detail in the upcoming sections. This motivated us to provide source domain information as a direct signal for each of the information snippets collected as well as a visualization of how all the collected snippets are distributed across the different domains, enabling users to easily assess source credibility and diversity.

\subsubsection{Information up-to-dateness}
There was a consensus among the participants that in order to make a correct decision, the evidence used must be up-to-date (11/15). Indeed, prior work also suggests that information currency is another crucial element contributing to its credibility, with the intuition that the older a piece of information is, the more obsolete it gets, which implies a lower level of trustworthiness \cite{alexander_web_1999,brandt_evaluating_1996,metzger_making_2007}. This is especially true in today's software development world, where languages and libraries are constantly being updated and older versions are quickly rendered obsolete by newer versions. For example, P6 was keen to stay on top of the state of the art of the \javascript frontend framework competition: \textit{``Is this speed comparison [between React, Angular, and Vue] up-to-date now that Angular 9 was just released?''} However, the above heuristic can be taken with a grain of salt by some participants, citing reasons that software that was updated a long time ago does not necessarily mean that it is obsolete. As P4 put it, \textit{``the last release of Haskell was like 10 years ago, but it's still the latest version, and I still use it all the time in my work.''} Nevertheless, we elect to provide users with direct access to at least the last updated timestamp information of each snippet that the author collected in an effort to help consumers assess up-to-dateness faster. In addition, the separate information about versions, as mentioned above, allows users to use whichever property is most relevant.

\subsubsection{Information popularity}
Echoing what has been reported in prior work that people seek social proof when evaluating information credibility \cite{seckler_trust_2015,metzger_making_2007}, participants (8/15) said that the popularity of information also plays an important role in its trustworthiness, with the general rule suggesting that the more people that stand behind a solution, the more trustworthy it is. For example, P9 said: \textit{``if there're a lot of other devs [who] also think this is a better idea, then I'm much more comfortable to use it.''} This is similar to the \textit{endorsement heuristic}  \cite{metzger_social_2010}, which suggests that people are inclined to perceive information and sources as credible if others do so too. This inspired us to directly present consumers with popularity signals (such as an answer's up-vote number on \stackoverflow, or the number of claps of an article on Medium.com) from where snippets are collected.

Also included in the endorsement heuristic is that people sometimes follow others' endorsements without much scrutiny of the site content or source itself \cite{metzger_social_2010}. However, some of our study participants suggest quite the opposite (7/15) --- they often put much more emphasis on source credibility over the popularity of specific information snippets from that source. For example, \textit{``in retrospect, if an answer is taken from \stackoverflow, I don't really care about its up-vote number or if it's the officially accepted one, I'll just trust it and use it''} (P3), or \textit{``I don't really look at how many people clapped over a Medium article, the fact that it's from Medium.com is usually good enough for me''} (P8). Though seemingly inconsistent with prior work, we do not claim that this is typical in the domain of programming --- one possible explanation is that websites like \stackoverflow by default rank the most up-voted posts at the very top with the specific intention to present the most popular information to readers.

\subsubsection{Information consistency}
In addition to source credibility, diversity, up-to-dateness, and popularity, a few participants (5/15) suggested that having more corroborating evidence implies that a piece of information is more trustworthy. For example, P6 said: \textit{``This [deep learning library comparison chart] claims that PyTorch is much easier to learn than Tensorflow, but I wonder if there're people suggesting otherwise? I kind of want to see at least one other expert that has experience with both and also says PyTorch is better.''} Prior research has also found that people will apply the \textit{consistency heuristic} to evaluate credibility, validating information by checking different websites to make sure that the information was consistent \cite{metzger_social_2010,meola_chucking_2004}. Meanwhile, consistency also implies the converse --- having contradicting evidence will undermine the trustworthiness of an existing piece of information.

\subsubsection{Author credibility}
Prior work has shown that the author's level of expertise impacts the credibility of information \cite{denning_wikipedia_2005,seckler_trust_2015}. This is especially significant in the domain of programming, where there is a substantial difference between novice and expert developers in their experience and ability to evaluate code and libraries \cite{begel_novice_2008}. For example, when shown with a comparison table on the topic of choosing a deep learning framework, P11 asked: \textit{``Does the author know what he's doing? I'd rather take advice from someone who's an expert rather than some random undergrad.''} However, participants (4/15) also reported that there is no easy way to tell the level of expertise of a table author or if that expertise matches with the topic of the table in the current \unakite system. 

Another factor that impacts the credibility of an author is if he or she is biased, possibly due to his or her affiliation or personal preferences --- for instance, P12 asked: \textit{``is the author saying all the nice things about Caffe \revision{[a deep learning framework]} because he has lots of experience with it or because he's biased?''} However, one participant also acknowledged that sometimes these ``biases'' may not be as negative as it sounds --- it could be an indication that an author is highly experienced with one particular option and therefore gives favorable evidence for it. To address the above concerns, prior research suggests that disclosing patterns of an author's past performance may be a good indication of his or her expertise as well as possible biases \cite{kittur_can_2008,shneiderman_designing_2000,suh_lifting_2008}. This motivated us to at least allow the author to provide a link to his or her GitHub profile, and \systemname will automatically compute and show relevant expertise metrics (contribution activities, most proficient programming languages, etc.) and affiliation information to the consumer.

\subsection{Thoroughness}
Another important facet when evaluating the appropriateness to reuse knowledge is thoroughness, which deals with the process and the amount of effort used when creating the knowledge, its coverage and scope, as well as any usable artifacts discovered or produced in the process.

\subsubsection{Research process and effort}
Prior work in sensemaking handoff recommends that when knowledge is handed-off from the author to the consumer, it should let the consumer be aware of the prior investigative process and insights \cite{paul_cosense:_2009,paul_sensemaking_2011,zhao_supporting_2018}, such as how much work has been done, and how mature the knowledge representation is \cite{sharma_sensemaking_2008,sharma_artifact_2009}. We also found relevant evidence from the interviews: three participants recalled similar experiences where they learned that the previous decision makers spent little time on exploring the decision space, and therefore the results were \textit{``too immature to be picked up and reused''} and \textit{``missing obvious criteria that you should definitely not leave out''}, and they ended up choosing to ignore those previous decisions and started from scratch to conduct their own research instead. This motivated us to automatically keep track of some of the authors' actions as they create tables using \unakite, such as the search queries used, the pages visited, the duration of their stay on each page and each query, etc. We then use these data to compute key statistics as well as timelines and visualize them to the consumers to help them better understand the author's research and exploration process.

P9 also envisioned that having a holistic understanding of the author's process would give her the ability to parse out the author's intention and focus (which may shift throughout the process, as discussed earlier), and therefore provide hints about what she needs to focus on next if she were to reuse this table as the basis for her own decision.

\subsubsection{Alternatives or competitors}
In addition to the process and effort, prior research recommends that knowledge and sensemaking results should also make apparent their coverage and scope \cite{denning_wikipedia_2005,metzger_making_2007}, for example, what alternatives have been considered, since not all options will necessarily appear in a \unakite table (especially when the author thinks one does not fit his or her particular needs and is therefore not worth further investigation). However, this does not necessarily imply that the option is inferior for the consumer. In our study, a few of our participants (6/15) were also interested in knowing what would those alternatives (or competitors) be and how they compare with the existing options before they could know if it is appropriate to reuse a table. For example, \textit{``I heard anecdotally that Svelte gives you much better performance than all these big (\javascript) frameworks [React, Angular, and Vue]. I should take a look at that before I decide. Or maybe there's again something else?''} (P14). This motivated us to take advantage of the Google Autocomplete API to automatically obtain commonly searched-for alternatives to the options that are already in the table, and present these alternatives to the consumers.

\subsubsection{Usable artifacts}
Lastly, participants (10/15) stressed the need for code examples and other usable artifacts from a decision, just as prior work reported that developers need help finding and reusing code examples \cite{brandt_example-centric_2010,brandt_two_2009,oney_codelets:_2012,ponzanelli_seahawk:_2013}. For example, P2 directly asked for code examples and the author's chosen option when presented with a decision table on various Java AST parsers: \textit{``[are there] any code snippets that I can immediately plug into mine and test? Or if you can tell me which is the one that the author used, I'll just try that one first.''} A few (3/15) participants also suggested that quickly trying out code examples to see if they work or not supersedes almost all other information needs. However, we do not claim this is typical, and later follow-up exchanges with these participants revealed that a vast majority of their current work is low-level detailed implementation, where making sure the code works is of paramount importance. Nevertheless, we implemented techniques to automatically extract code blocks from various snippets and present them to consumers. In addition, we also detect authors' copy events in the browser, and use those as the basis for a heuristic to tell which option the author chose for the decision.

\subsection{Summary}
We found that when evaluating the appropriateness to reuse a piece of knowledge, one should not only assess its trustworthiness (as the majority of the prior research has focused on), but also check for its context and thoroughness. However, no previous system has made significant attempts to address developers' specific information needs with regard to all three of these facets, or to extract appropriateness properties from the original content and present them to the consumer of the knowledge to facilitate reuse. In addition, this process should not put much burden on either the author or the consumer \cite{liu_unakite:_2019,van_de_vanter_documentary_2002} by requiring them to manually locate those appropriateness properties, suggesting the need for largely automatic mechanisms.

\begin{figure}[t]
\centering
	\includegraphics[width=\linewidth]{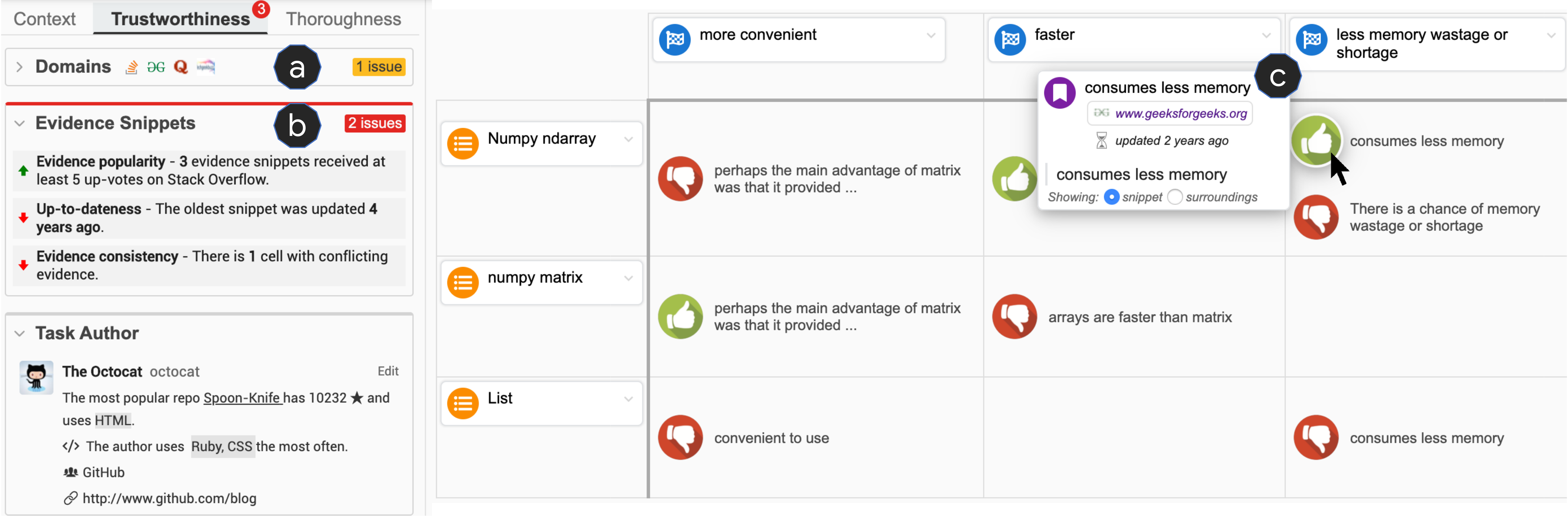}
	\caption{On \systemname startup, none of the groups are activated to keep the \unakite table on the right clean and concise. Groups can also be collapsed to keep the sidebar interface clean (such as (a)). Mousing over each snippet in the table will only show the exact content that an author captured by default (c), the same as the original \unakite system, rather than the automatically captured \textit{context snapshots}. Only after a user activates some groups in the \systemname sidebar (by clicking on their titles) will the corresponding additional metadata appear on the snippets in the table, as shown in Figure \ref{fig:trust-panel}.}
	\label{fig:non-activated}
\end{figure}

%
%
%
%
%
%
%
\section{\systemname Design and Implementation} 
\label{sec:design-and-implementation}

Based on the findings in our interviews and the framework, we built a prototype system called \systemname to visualize properties and signals of the appropriateness to reuse for the consumers of a decision. 

\subsection{Core Design Process and Rationale}
We first consulted the interview data and brainstormed the various signals and properties that would theoretically address each of the information need listed in Table \ref{tab:framework-table}, column 2. Some information needs can be directly addressed by obvious signals, such as surfacing the domain names of the source web pages to consumers so that they know where the information in the table were collected from and if those sources are credible. For information needs that would require explicit effort from the table author to provide, such as the goal of a decision, we also consulted prior literature as well as brainstormed about potential indirect signals that can be used by consumers to infer those needs. For example, search queries are useful for inferring task goals and contexts of an author \cite{bharat_searchpad_2000,morris_searchtogether:_2007,paul_cosense:_2009,sharma_artifact_2009}.

In order to obtain these signals, we then built tracking techniques to automatically keep track of the author's activities in the browser while searching and browsing during the creation of a \unakite table. Many of these tracking and extraction techniques use heuristics that are based on the current design of websites that developers most often use, such as extracting the number of up-votes for an answer on a \stackoverflow page. These are meant as a proof-of-concept, and more elaborate and crowd-sourced extraction techniques could be added in the future. 

We then set out to design a visualization that presents the consumers with these signals and properties. During our exploration of the design space, we struggled with a fundamental tension between consumers' awareness of all the signals and consumers' limited attention bandwidth. In our initial prototypes, we placed all the signals (approximately 15) in a scrollable vertical list to the left of the original \unakite table. Users would also be able to hide a signal if it was not relevant. We hoped to make the users aware of all the signals that \systemname can provide and give them complete freedom to explore them as they wish. Another rationale for this design was that users would be able to use a combination of signals to fulfill a single information need, for example, both the search queries and the pages visited will help indicate the author's research process and effort, as evidenced by the formative interviews. However, by implementing and testing these design probes with a convenience sample of 8 developers, we realized that having \textit{``everything all at once''} can be overwhelming to the consumers, and they would prefer to just examine one facet at a time and tune out the ``noise'' (signals that are irrelevant to the facet currently being examined). In addition, we found that there was a disconnect between the signals we showed in the list on the left and the actual content in the table on the right, causing consumers the additional mental burden of trying to match them up. Showing the signals in context along with the various information snippets in the table seemed to be a much better design to address this problem.

These findings guided us towards a hierarchical visualization design of \systemname's consumer-facing user interface: to structure these properties and guide the consumers through their evaluation process, we designed \systemname as a sidebar to a \unakite table. \systemname's sidebar contains three tabbed overview panels for the three facets in the aforementioned framework (Figure \ref{fig:trust-panel}-a). Each overview panel provides multiple \textit{groups} (e.g., Figure \ref{fig:trust-panel}-b,c,d) of appropriateness properties to directly address consumers' information needs as summarized in the framework. In addition, by activating one or more of the groups (by clicking on their titles in the sidebar), consumers will be able to view additional information specific to each snippet in the table. For example, Figure \ref{fig:non-activated} shows a state where none of the groups are activated. After activating the Domains group and Evidence Snippets group, consumers will be able to see for each snippet: where it originated (Figure \ref{fig:trust-panel}-g1), how popular it is (Figure \ref{fig:trust-panel}-g2,3), and how old it is (Figure \ref{fig:trust-panel}-g4). This is designed to provide consumers \revision{with} a high-level overview of each of the facets of reuse as well as the ability to dive into the parts of interest, as recommended by Shneiderman \cite{shneiderman_eyes_1996}. It is also inspired by the \textit{lens} interaction \cite{bier_toolglass_1993,chang_searchlens_2019} where the same table content is addressed from three different perspectives.

Like \unakite, \systemname consists of an extension to the Chrome Web browser and a web application. \systemname's Chrome extension implements the aforementioned new tracking techniques on top of the \unakite Chrome extension. The \systemname web application is implemented in HTML, \javascript, and CSS, using the React \javascript library \cite{facebook_react_2018} as the primary frontend UI development framework and Google's Firebase on the Google Cloud for data management and synchronization as well as user authentication. 

We now discuss how the different features in \systemname support the three facets listed in the previous framework, and how they are implemented.

\begin{figure}[t]
\centering
	\includegraphics[width=\linewidth]{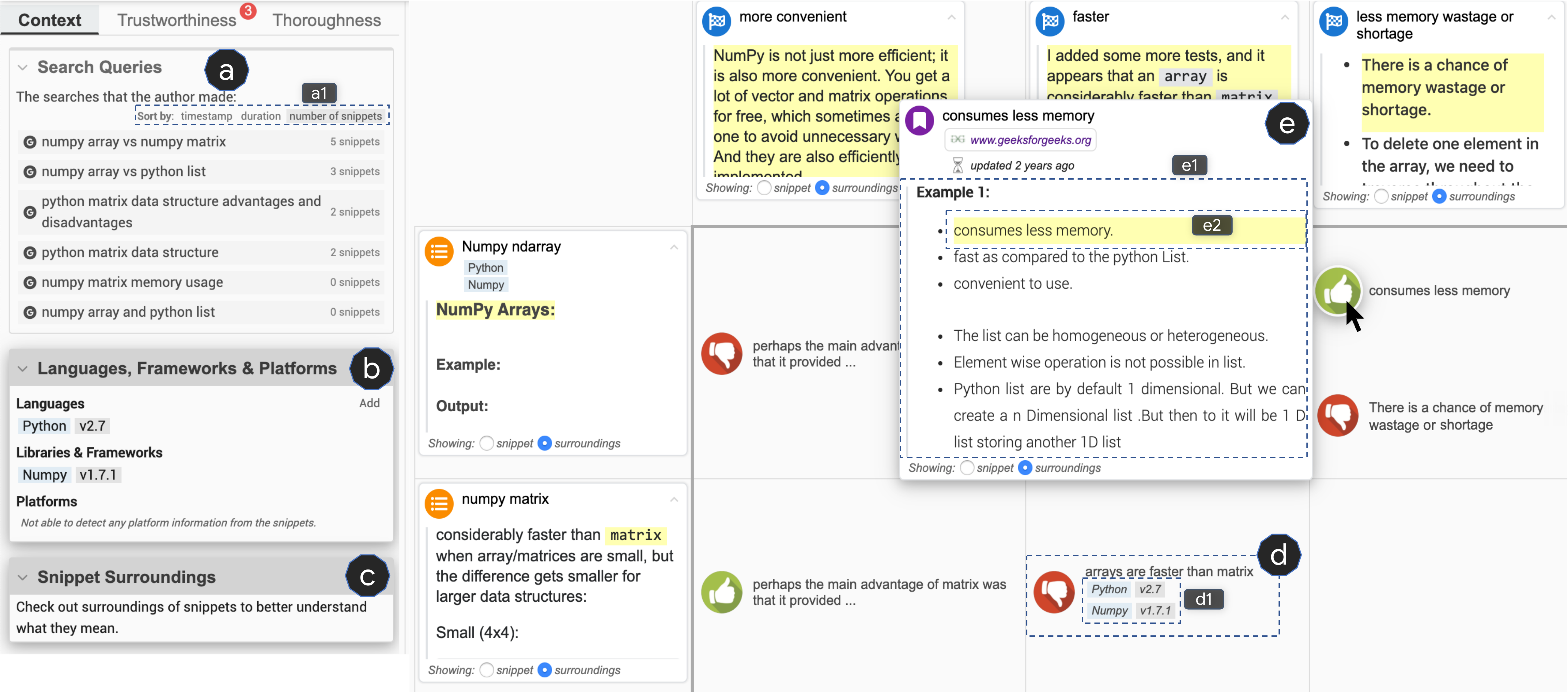}
	\caption{\systemname's \textit{Context} panel. Consumers are able to check the search queries (a) that the author used to understand his or her goal, examine the languages, frameworks, platforms, and their versions of the snippets (b, d1), and view the surroundings of a snippet through the automatically captured context snapshots (e1).}
	\label{fig:context-panel}
\end{figure}

\subsection{Context}
\subsubsection{Capturing goals with search queries}
First of all, \systemname automatically keeps track of authors' search queries used in \unakite tasks as well as the duration of time they spent on each and the number of information snippets they collected. The duration information is approximated by comparing the timestamp when the next query is issued to that of the current one. It also automatically leaves out any idle time (i.e., time where there is no activities detected in the browser, by monitoring mouse movements, keyboard input, etc.) that are longer than a certain threshold to make the duration approximation more accurate. The idle threshold was empirically tuned to be 8 seconds based on data obtained through pilot testing, and can be flexibly adjusted in the future. For consumers, \systemname visualizes these search queries as a list (Figure \ref{fig:context-panel}-a) to help consumers understand the goals of the task author. They can use the sorting mechanisms at the top (Figure \ref{fig:context-panel}-a1) to sort the search queries by chronological order, by duration, or by the number of information snippets yielded from each (which is the default sorting order, where ties are broken by ascending chronological order). 

There are several advantages of using search queries as a representation of an author's goals. First, they are direct translations of what an author thinks and intends to do to satisfy their information need \cite{sadikov_clustering_2010} --- for example, issuing the query ``numpy matrix vs list'' implies that the author would like to find out the differences between the two options. Second, unlike the original \unakite where an author sets the single task goal (as the name of a task) at the beginning, keeping track of all of the search queries (in temporal order) captures not only the author's original goal (which usually is the first query based on pilot study data) but also the evolving nature of the goal (as identified in the formative interviews). Third, the number of snippets yielded from each query serves as an approximation of an author's effort spent on that particular part of the task, which informs consumers of the author's focus throughout the decision making process.

\subsubsection{Contextualizing information with automatic context snapshots}
To help consumers contextualize and understand the meanings of options, criteria, and evidence in \unakite (identified as one of participants' frustrations), \systemname introduces the idea of automatically keeping a snapshot of the surroundings of a piece of content called \textit{context snapshot} (inspired by \cite{hu_screentrack_2020}) as an author collects information snippets. \systemname uses \unakite's \textit{snapshot} feature, where website content can be captured and preserved with its original styling, including the rich, interactive multimedia objects supported by HTML. The bounds of the surroundings are by default defined as the main content (\systemname automatically tries to exclude any advertisements and other forms of injected content on a website) in the visible area of a web page in the browser window. In addition, due to the popularity and importance of \stackoverflow in the domain of programming, we specifically optimized this feature to include not only the particular answer block an author collects information from but also the original question block regardless of whether they are within the bounds, which provides consumers with extra context information. Similar optimizations for other popular developer sites, such as the official documentation, could be added in the future. On the consumer side, by clicking on the title of the \textit{Snippet Surroundings} group (Figure \ref{fig:context-panel}-c) in the \systemname sidebar, consumers will be able to view and scroll through the surroundings for each snippet (Figure \ref{fig:context-panel}-e1), with the content that the author specifically collected highlighted in yellow (Figure \ref{fig:context-panel}-e2).

This feature offers several benefits to both the authors and the consumers. The surrounding of a snippet is highly likely to include explicit explanations (such as screenshots, code examples, and execution results) that can help consumers understand exactly what a snippet means. For example, the \textit{Python Lists VS Numpy Arrays} article \cite{noauthor_python_2020} where a criterion snippet ``more efficient'' was scooped from, also gives examples of how the two data structures allocate memory blocks under the hood, suggesting that the author actually meant ``more \textbf{memory} efficient'' rather than ``more \textbf{time} efficient''. Unlike in \unakite, where an author needs to specifically include that entire paragraph when creating a snippet and then manually change the title of the snippet into ``more memory efficient'' (which may disrupt the workflow), \systemname will automatically capture that helpful paragraph into the snippet's context snapshot. During the evaluation of context, consumers will be able to directly view a snippet in its surroundings through its context snapshot without frequent switches to the corresponding original web page to find where the content where the snippet was taken from (which is exactly what participants reported doing in the formative study interviews).

\subsubsection{Detecting languages, frameworks, and their versions}
\systemname tries to automatically detect the languages, frameworks, platforms, and their versions used in the snippets to directly address consumers' information needs. To ground this feature, we picked the top 10 of each of the most popular languages, frameworks, and platforms from the \revision{2020 \stackoverflow developer survey \cite{noauthor_stack_2020}} and built \textit{detectors} for them. The detectors for a language (or a framework, platform, etc.) is implemented as a set of manually devised keywords (e.g., language statements, special variables, file extensions, etc.) that can uniquely identify the usage or presence of that language. For example, ``\texttt{es7}'', ``\texttt{console.log}'', ``\texttt{setTimeout}'', etc. can be used to identify \textit{\javascript}, and ``\texttt{useState}'', ``\texttt{componentDidMount}'', ``\texttt{findDOMNode}'', etc. and be used to identify the \textit{React} library. Keywords that can cause ambiguities are specifically avoided, such as ``\texttt{\$}'' (the dollar sign) is simultaneously a way to refer to variables in \textit{PHP} and a shortcut for \textit{jQuery}. \systemname then automatically tries to find these detectors through optimized string matching in a snippet upon its collection. If there is no hit within the snippet content, \systemname will make a second attempt with the content of the snippet's parent web page. Subsequently, \systemname uses regular expressions to find version numbers in the vicinity of detected languages, frameworks, and platforms (e.g., ``\texttt{Angular \textbf{9}}'', ``\texttt{Python \textbf{3.5}}'', ``\texttt{React \textbf{16.13.1}}'', etc.) or in the web page's URL (e.g., Java SDK version numbers are encoded in the URL of its official documentation website). In an informal evaluation using materials containing only the currently supported languages, this mechanism was able to successfully extract language information 100\% of the time and correctly identify the version information 96\% of the time. In the future, one might imagine \systemname pulling detectors from open-source detector repositories built, verified, and maintained by the community, which can improve their quality, precision, and recall, or at the very least, letting authors add or correct wrongly detected versions. On the consumer side, this detected information is then presented directly on the corresponding snippet cards in the table (Figure \ref{fig:context-panel}-d) as well as aggregated in the \textit{Languages, Frameworks, and Platforms} group (Figure \ref{fig:context-panel}-b). 

Directly surfacing these language, framework, and platform version entries to consumers will help them quickly understand the technologies used in the task as well as the specific versions each snippet uses at a glance, to support comparing those with their own situation. For example, one developer would be easily able to figure out that the example code collected by the other developer uses Python 2.7 and therefore does not match with \revision{his or her} own environment, which uses Python 3.5.

\begin{figure}[t]
\centering
	\includegraphics[width=0.4\linewidth]{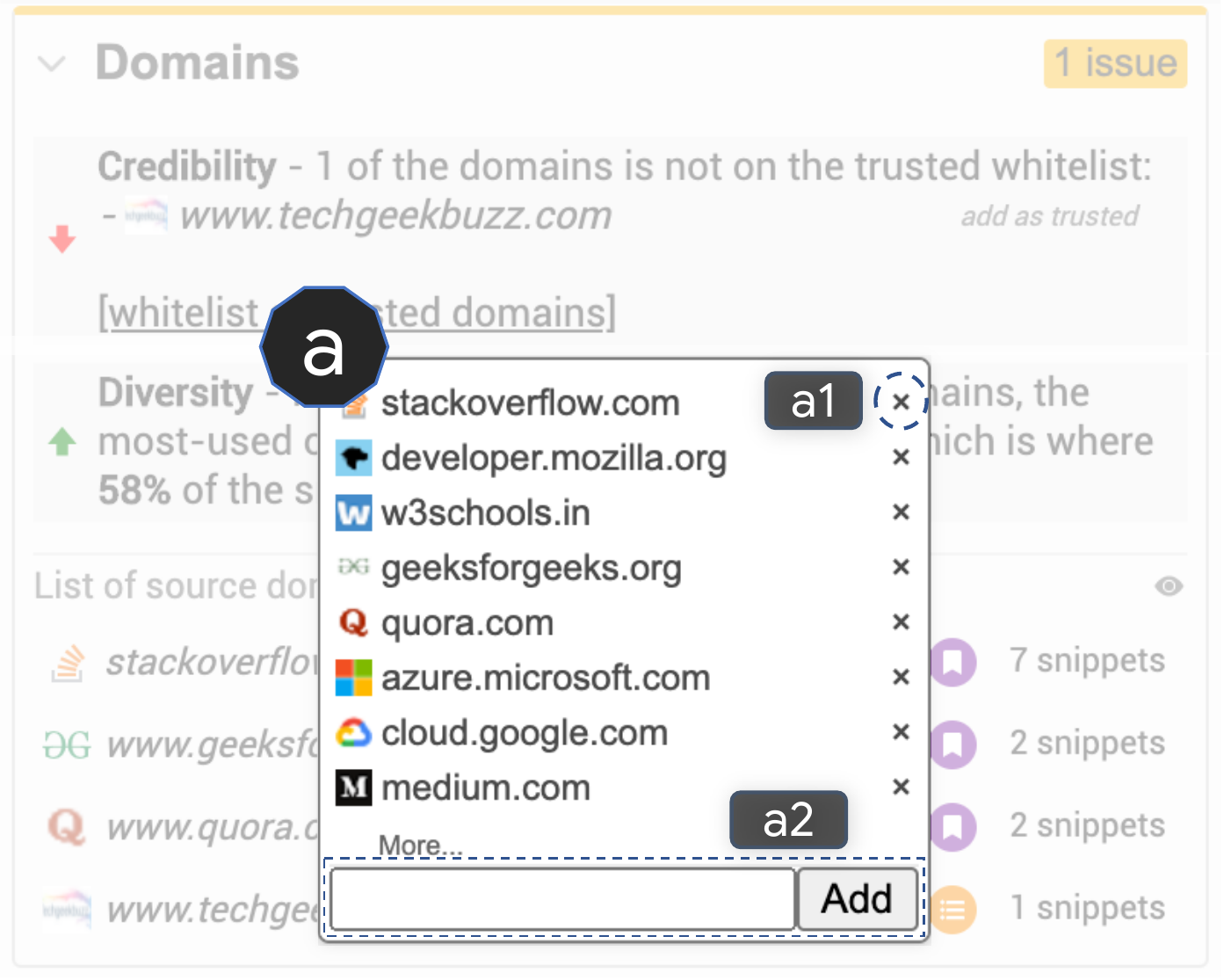}
	\caption{The \textit{trusted domains whitelist}. Consumers are able to remove a certain domain from the whitelist (a1) or add new ones (a2).}
	\label{fig:trusted-domain-whitelist}
\end{figure}

\subsection{Trustworthiness}
To help consumers evaluate the trustworthiness of a table, \systemname provides visualizations of various properties that directly address their information needs listed in the framework (e.g., source credibility, information popularity, etc.). Prior work has suggested that surfacing issues or problems that could cause distrust is an effective way to alert and guide users' attention during credibility evaluations \cite{metzger_making_2007}. Therefore, in addition to visualizing the trustworthiness properties, we remind users of potential issues that could negatively impact a table's trustworthiness by marking them with a red downward arrow (Figure \ref{fig:trust-panel}-b2,c3,c4). The count of the number of issues is shown in a colored badge on the top-right corner of the Trustworthiness panel (Figure \ref{fig:trust-panel}-a1), with one issue having a yellow color, and more than one issue having a red color (these user-adjustable levels were empirically determined). Future development will explore more sophisticated weighting of the issues beyond counting them equally.

\subsubsection{Visualizing source credibility and diversity}
As shown in Figure \ref{fig:trust-panel}-b, \systemname visualizes the distribution of the snippets across different domains (websites) (Figure \ref{fig:trust-panel}-b5), giving consumers a high-level overview of the provenance of the information in the table. In addition, each snippet in the table is also marked with its domain (Figure \ref{fig:trust-panel}-g1), giving consumers a detailed understanding of where each snippet originated.

\systemname also alerts consumers of potential untrusted domains by checking the presence of each domain on a user-defined \textit{trusted domains whitelist}, and flags the ones that are not on the list. For example, a consumer will be able to immediately notice that one of the websites that the author used to collect evidence, \texttt{techgeekbuzz.com}, is not on \revision{his or her} own trusted domains whitelist (Figure \ref{fig:trust-panel}-b2). Currently, the default whitelist was generated by mining and aggregating the websites that 5 full-stack developers (who work for different technology companies and routinely use a variety of languages and technology stacks) visited from their browsing history. We then had them each annotate the websites as either ``credible'' or ``not credible'', and removed the ones that they did not all agree upon. This resulted in 25 domains that are considered ``credible'', including community Q\&A sites like \texttt{stackoverflow.com}, official documentation sites like \texttt{angular.io}, and blog sites like \texttt{medium.com}. Domains that sometimes contain non-objective and low-quality information are rejected, such as \texttt{reddit.com}. We by no means claim this is complete nor that it applies to everybody --- instead, it serves as a starting point and the consumers are able to add and remove items themselves (Figure \ref{fig:trusted-domain-whitelist}-a1,a2). They can also use the ``add as trusted'' button  (Figure \ref{fig:trust-panel}-b3) to add a flagged website to the whitelist so that any future information originating from that website will not be considered as an issue. In the future, one can imagine taking advantage of a larger consumer base and automatically marking websites as trusted if a majority of the consumers have it on their whitelist. We also expect to periodically update the default whitelist over time, as new programming technologies are created and become popular in the future.

To help with the evaluation of source \textit{diversity}, \systemname also alerts consumers when there is only limited sources used to construct a table. Currently, \systemname considers that there is an issue in terms of source diversity if all of the information comes from one single source (reported by participants in the formative studies as the worst scenario). If that is the case, the green upward arrow for source diversity in Figure \ref{fig:trust-panel}-b4 will become a red downward arrow, reminding consumers that it is an issue. However, this threshold can be set by individual consumers, which would then apply to all future table evaluations they perform. Similar to source credibility issues, this can also be resolved or dismissed by individual consumers if they do not think it is problematic.

\subsubsection{Examining evidence trustworthiness}
Consumers will be able to get information about the popularity, up-to-dateness, and the consistencies of the evidence by activating the \textit{Evidence Snippets} group (Figure \ref{fig:trust-panel}-c).

Each snippet in the table will be marked with signals showing its popularity depending on the websites and pages that it originates from. For example, if a snippet is collected from a \stackoverflow answer post, \systemname will automatically extract and show the up-vote number of that post (Figure \ref{fig:trust-panel}-g2) as well as if that answer is the officially accepted answer (Figure \ref{fig:trust-panel}-g3). If a snippet is collected from a Medium.com article, \systemname will show the number of claps that article had at the time of collection. We designed this feature to closely fit developers' current ways of evaluating popularity, as reported in the formative studies. \systemname will also display an alert in the Evidence Snippets group if some of the snippets in the table have particularly low popularity, such as \revision{down-votes} on \stackoverflow. As with the other kinds of detectors, we envision these being augmented over time based on where developers are mostly getting their information from.

Unlike the original \unakite, which only showed \textit{when} information was collected (reported as \textit{``not exactly helpful''} by participants in the formative interviews), each snippet in the table will be marked by \systemname with the timestamp of when its parent webpage (or answer post if it is from \stackoverflow) was last updated (Figure \ref{fig:trust-panel}-g4). \systemname uses a combination of techniques to extract the last updated timestamp information, including using regular expressions to look for date strings in website source code and taking advantage of the \javascript \texttt{document.lastModified} variable (only when the website is static). This serves as a direct measurement of the age of information, and gives consumers an idea of how old the information is. Our study participants also mentioned that they often had trouble quickly locating when articles or blogs are updated online as these timestamps are often displayed in less salient font styles or not visible at all. In addition, \systemname will flag snippets that are older than 3 years as a potential issue in the Evidence Snippets group (Figure \ref{fig:trust-panel}-c3), which, similar to other issues, can be manually adjusted or dismissed by the consumer.

Finally, \systemname provides initial support for information consistency by informing consumers if there are corroborating or conflicting evidence snippets in a table cell (e.g., there are simultaneously both thumbs-up and thumbs-down ratings for ``\texttt{numpy ndarray}'' causing ``less memory wastage or shortage'') (Figure \ref{fig:trust-panel}-c4). The culprit table cells with conflicting evidence will be highlighted by mousing over the issue in the Evidence Snippets group, addressing concerns from participants in the formative studies about how such contradictions could be overlooked once a table gets larger with more evidence ratings.

\subsubsection{Surfacing properties about author credibility}
\systemname provides consumers with help in evaluating author credibility by allowing authors to manually provide information about themselves. In the current implementation, a table author can input a link to their GitHub profile, and \systemname will automatically present the author's name, numbers of stars on the most popular code repositories \revision{he or she owns}, most used programming languages, affiliation, and a link to his or her GitHub profile page in the \textit{Task Author} group (Figure \ref{fig:trust-panel}-d). We opted to let authors voluntarily provide this information in order to give them the option to protect their privacy and identity. In the future, we will work on mechanisms to automatically perform author modeling in a privacy-preserving way --- one idea is to analyze the topics of \stackoverflow questions and coding forums that an author frequently visits to infer his or her expertise. We will also provide an option for authors to provide certain information to consumers anonymously.

\begin{figure}[t]
\centering
	\includegraphics[width=\linewidth]{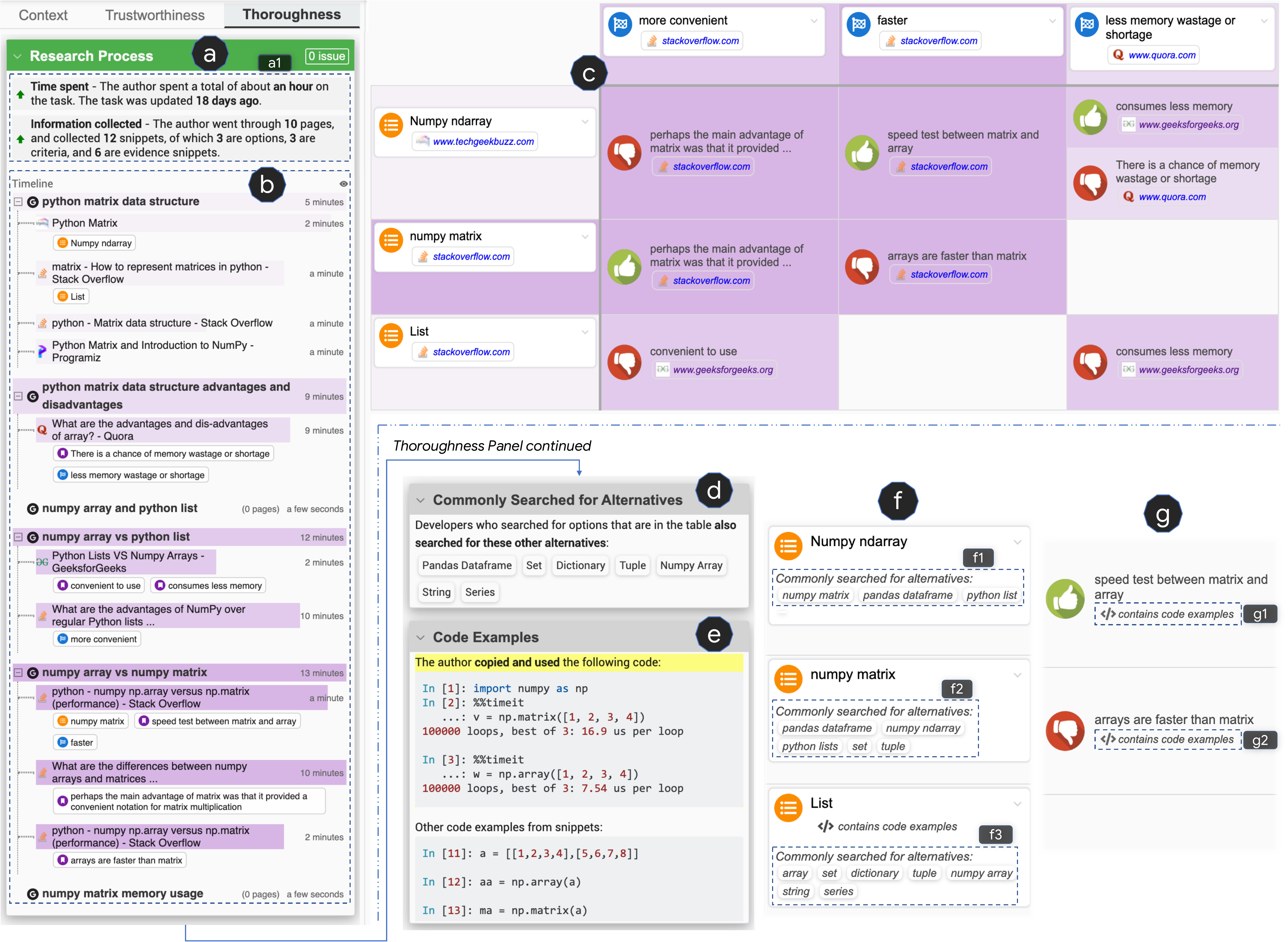}
	\caption{\systemname's \textit{Thoroughness} panel. Consumers are able to understand the author's research process (a) with the help of the timeline view (b) (a lighter violet means older chronologically), check commonly searched for alternatives to the existing options (d, f1, f2, f3), and check the code examples in the snippets (e).}
	\label{fig:thoroughness-panel}
\end{figure}

\subsection{Thoroughness}

\subsubsection{Understanding the research process}
In order to provide consumers with a clear understanding of an author's research and exploration process, \systemname automatically keeps track of several of the author's activities in the background --- in addition to the search query tracking discussed earlier, \systemname also automatically records the web pages visited, as well as the time spent, progress made (approximated by tracking the percentage of a page that has been scrolled into the visible browser window using \javascript's \texttt{window.onscroll} event), and the number of information snippets collected on each of the web pages. 

With these activity data, \systemname computes the duration of time the author spent working on a task, the length of time since the task was last updated by the author, and the numbers of options, criteria, and evidence snippets that the author collected (Figure \ref{fig:thoroughness-panel}-a1).

In addition, \systemname visualizes the activity information on a timeline view (Figure \ref{fig:thoroughness-panel}-b), which provides an integrated chronological representation of the author's entire research and exploration process during a task. The timeline view is organized with two levels of hierarchies: first by the search queries, and then by the pages that are visited during a particular search. The timeline view is color-coded by different shades of a violet color, with increasing intensity indicating the chronological order (a lighter violet means older). The same color scheme is also applied to the background of the table cells (Figure \ref{fig:thoroughness-panel}-c) when the \textit{Research Process} group is activated. The timeline view is also interactive, mousing over a search query or a page will highlight its corresponding information snippets in the table, together with the colored background, giving consumers an understanding of how the table was constructed chronologically.

\subsubsection{Suggesting alternatives}
Another way for \systemname to help with the thoroughness evaluation is to provide consumers with \textit{commonly searched for alternatives} to each option (Figure \ref{fig:thoroughness-panel}-f1,f2,f3). For every option in the table, \systemname will automatically obtain the potential alternatives to that option by making Google search queries in the form of ``[option\_name] vs'' or ``[option\_name] versus'' and obtaining a list of top 10 auto-complete candidates using the Google Autocomplete API. This will then be transformed into the \textit{alternatives list} for the corresponding option by extracting and cleaning the part after ``vs'' or ``versus'' for each auto-complete candidate, followed by aggregating and removing duplicates. The results are presented in the \textit{Commonly Searched for Alternatives} group (Figure \ref{fig:thoroughness-panel}-d). These alternative lists are generated on the spot every time a table is being reviewed, making sure that \systemname always presents the latest information. 

This approach offers several benefits to the consumers of the table. First, it offers insights into the popularity of the existing options in the table --- if an option (such as ``React'') appears in all other options' alternatives lists (such as for ``Angular'' and ``Vue''), it suggests that this option has a high popularity. Second, it provides consumers with an understanding of the coverage of the author's research process as well as guidance on potential new opportunities to explore next --- if an item (such as ``pandas dataframe'' in Figure \ref{fig:thoroughness-panel}-d) frequently appears in the existing options' alternatives lists (and therefore will rank higher in the aggregated list in the Commonly Searched for Alternatives group), it suggests that this item might have been overlooked by the author initially, or it might not have been available back when the table was made, and the consumers can focus their investigative effort on it next before deciding whether to reuse this table. This feature could help authors as well, offering real-time reminders of the coverage of their research process and possible new options to consider as they are making decisions.

\subsubsection{Presenting usable artifacts}
Finally, \systemname automatically detects and extracts any code examples included in the collected snippets and presents them in the \textit{Code Examples} group under the Thoroughness panel (Figure \ref{fig:thoroughness-panel}-e). This provides consumers the opportunity to directly examine and try out any code examples involved first without diving deeper into the table. In addition, when the Code Examples group is activated, a \textit{``contains code examples''} badge (Figure \ref{fig:thoroughness-panel}-g1,g2)  will appear on snippets that contain code examples, helping consumers quickly locate potential code examples for a particular option or criterion in the table.

%
%
%
%
%
%

%
%
%
%
%
%
\section{Evaluation}
\label{sec:eval}
We conducted a lab study to evaluate the effectiveness of the framework and the prototype \systemname system in helping developers evaluate the appropriateness of reusing decisions.
%

\newcommand{\placeholderNum}{{\color{red}x}\xspace}
\subsection{Experiment Design}
\subsubsection{Participants}
We recruited 20 participants (13 male, 7 female) aged 22-37 ($\mu = 26.95$, $\sigma = 3.81$) years old through emails and social media. The participants were required to be 18 or older, fluent in English, and experienced in programming. Participants on average had 8.3 ($\sigma = 3.3$) years of programming experience, with 11 of them currently working or having worked as a professional developer and the rest having programming experience in universities.

\subsubsection{Procedure}
Participants were presented with 3 tasks in random order. The topics of the tasks were: \textit{(a) choosing a python data structure to represent matrix-like data} (referred to as \textit{Python} from here on), \textit{(b) choosing a deep learning framework to build neural networks} (referred to as \textit{Deep} from here on), and \textit{(c) choosing a cloud computing service to build a video-streaming application} (referred to as \textit{Cloud} from here on). For each task, participants were told what to pretend their background and context was, and they needed to read a table and answer questions about: (1) how much do they think the table is relevant to their given background and context; (2) how much do they trust the content of the table; and (3) to what extent do they think the research effort put into making the table is thorough. Participants were required to list out specific reasons to justify their evaluations.


The study was a between-subjects design, where participants were randomly assigned to either the \systemname condition or the \unakite (control) condition. In the \systemname condition, participants had full access to all the \systemname features described above (along with the table produced by \unakite), while in the \unakite condition, these new features were turned off, so the participants saw only the table, and snippets in the table only showed their titles, contents, timestamps of collection, and links to their original web pages. We imposed a 10-minute limit per task to keep participants from getting caught up in one of the tasks. However, participants were instructed to inform the researcher when they thought they had finished the task or felt like they could make no further progress. 

We chose \unakite as the control condition as opposed to raw (and textual) comparison tables online to make sure both conditions had a similar user interface to work with. It also makes the comparison between conditions more realistic --- since the original \unakite is already keeping track of where snippets are collected, participants in the \unakite condition would have the ability to go back to the source to examine the appropriateness signals (such as up-vote numbers, last-updated timestamp, etc.) if they wanted to.

Each study session started by obtaining the proper consent and having the participant fill out a demographic survey. Participants in the \unakite condition were given a 10-minute tutorial showcasing the various features of the \unakite web application as well as a practice task on the topic of ``choosing a \javascript frontend framework'' before starting. Those in the \systemname condition were given a same-length tutorial as well as the same practice task but in \systemname instead. At the end of the study, the participant was invited to fill out a questionnaire focusing on the experience of using either \systemname or \unakite. We asked questions on the usability of the system they used in their respective conditions, the usefulness of such tables generated by the system, their opinions of the different features of the system, their willingness to author tables using the system to keep track of their decisions, their concerns about privacy if they were to author tables, as well as their familiarity with the topic of the three tasks used in the study. Finally, we ended the session with an informal interview on any additional thoughts they had about the system they used. Each study session took about 60 minutes per participant and was done remotely using the Zoom video-conferencing application. All participants were compensated \$15 for their time.

\newcommand{\statsSig}[1]{#1*}
\begin{table}[t]
\centering

\begin{subtable}{1\textwidth}
\resizebox{1\textwidth}{!}{%
\begin{tabular}{l | l | p{16mm} | p{16mm}p{23mm}p{21mm}p{16mm} | p{16mm} p{20mm} p{20mm} }
    \toprule
    & 
    Time &
    $n_\textbf{Total}$ &
    $n_\textbf{Valid}$ for\newline Context & 
    $n_\textbf{Valid}$ for\newline Trustworthiness & 
    $n_\textbf{Valid}$ for\newline Thoroughness &
    $n_\textbf{Valid}$ &
    $n_\textbf{High Quality}$ &
    Precision &
    Recall
    \\
    \midrule
    
    \unakite &
    \statsSig{484.2 (37.8)} & 
    \statsSig{5.20 (0.92)} &
    1.50 (0.53) &
    \statsSig{1.30 (0.48)} &
    \statsSig{1.20 (0.42)} &
    \statsSig{4.00 (0.67)} &
    \statsSig{2.90 (0.57)} &
    \statsSig{55.7\% (4.9\%)} &
    \statsSig{24.2\% (4.7\%)}
     \\
    
    \systemname &
    \statsSig{328.2 (48.1)} & 
    \statsSig{7.90 (1.91)} &
    1.50 (0.53) &
    \statsSig{3.20 (0.79)} & 
    \statsSig{2.70 (0.82)} &
    \statsSig{7.40 (1.51)} &
    \statsSig{7.10 (1.45)} &
    \statsSig{90.1\% (6.8\%)} &
    \statsSig{59.2\% (12.1\%)}
    \\
    \bottomrule
\end{tabular}%
}
\caption{Python ($n_\textbf{Ref. High Quality} = 12$)}
\end{subtable}

\begin{subtable}{1\textwidth}
\resizebox{1\textwidth}{!}{%
\begin{tabular}{l | l | p{16mm} | p{16mm}p{23mm}p{21mm}p{16mm} | p{16mm} p{20mm} p{20mm} }
     \toprule
    & 
     Time &
    $n_\textbf{Total}$ &
    $n_\textbf{Valid}$ for\newline Context & 
    $n_\textbf{Valid}$ for\newline Trustworthiness & 
    $n_\textbf{Valid}$ for\newline Thoroughness &
    $n_\textbf{Valid}$ &
    $n_\textbf{High Quality}$ &
    Precision &
    Recall
    \\
    \midrule
    
    \unakite &
    \statsSig{393.4 (50.9)} &
    \statsSig{5.70 (1.06)} &
    1.70 (0.48) &
    \statsSig{1.60 (0.70)} &
    \statsSig{1.40 (0.52)} &
    \statsSig{4.70 (0.82)} &
    \statsSig{3.20 (0.92)} & 
    \statsSig{56.1\% (12.4\%)} &
    \statsSig{29.1\% (8.3\%)}    
     \\
    
    \systemname &
    \statsSig{276.2 (68.3)} &
    \statsSig{7.80 (1.87)} &
    1.70 (0.67) &
    \statsSig{3.00 (1.15)} &
    \statsSig{2.60 (0.70)} &
    \statsSig{7.30 (1.83)} &
    \statsSig{6.90 (1.97)} & 
    \statsSig{88.1\% (9.7\%)} &
    \statsSig{64.5\% (17.4\%)}   
    \\
    \bottomrule
\end{tabular}%
}
\caption{Deep ($n_\textbf{Ref. High Quality} = 11$)}
\end{subtable}

\begin{subtable}{1\textwidth}
\resizebox{1\textwidth}{!}{%
\begin{tabular}{l | l | p{16mm} | p{16mm}p{23mm}p{21mm}p{16mm} | p{16mm} p{20mm} p{20mm} }
     \toprule
    & 
     Time &
    $n_\textbf{Total}$ &
    $n_\textbf{Valid}$ for\newline Context & 
    $n_\textbf{Valid}$ for\newline Trustworthiness & 
    $n_\textbf{Valid}$ for\newline Thoroughness &
    $n_\textbf{Valid}$ &
    $n_\textbf{High Quality}$ &
    Precision &
    Recall
    \\
    \midrule
    
    \unakite &
    \statsSig{420.4 (58.9)} &
    \statsSig{6.20 (1.03)} &
    \statsSig{1.40 (0.51)} &
    \statsSig{1.90 (0.74)} &
    \statsSig{1.50 (0.53)} &
    \statsSig{4.80 (1.14)} &   
    \statsSig{3.60 (0.97)} &
    \statsSig{58.5\% (15.2\%)} &
    \statsSig{30.0\% (8.1\%)}
     \\
    
    \systemname &
    \statsSig{271.8 (35.3)} &
    \statsSig{9.60 (2.37)} &
    \statsSig{2.60 (0.84)} &
    \statsSig{3.80 (0.92)} &
    \statsSig{2.60 (0.70)} &
    \statsSig{9.00 (2.00)} &   
    \statsSig{7.90 (1.45)} &
    \statsSig{83.8\% (8.5\%)} &
    \statsSig{65.8\% (12.1\%)}
    \\
    \bottomrule
\end{tabular}%
}
\caption{Cloud ($n_\textbf{Ref. High Quality} = 12$)}
\end{subtable}

\caption{Lab study results. The numbers of gold standard high quality reasons for each task, $n_\textbf{Ref. High Quality}$, are listed in their respective captions. We report the mean and standard deviation for: (1) the \textbf{time} in seconds taken to finish a task; (2) the total number of reasons participants came up with, $n_\textbf{Total}$; (3) the number of valid reasons, $n_\textbf{Valid}$; (4) the number of high quality reasons, $n_\textbf{High Quality}$; (5) the precision of high quality reasons, calculated as $n_\textbf{High Quality} / n_\textbf{Total}$; (6) as well as the recall of high quality reasons, calculated as $n_\textbf{High Quality} / n_\textbf{Ref. High Quality}$. Statistically significant differences (p < 0.05) through t-tests are marked with an *.}
\vspace{-8mm}
\label{tab:study-results-table}
\end{table}

\subsection{Quantitative Results}

All participants were able to complete all of the tasks in both conditions, and none of them went over the pre-imposed time limit. 

The results show that the participants in the \systemname condition took significantly \textit{less time} to finish compared to the \unakite condition for all three tasks, as shown in Table \ref{tab:study-results-table}. Across all three tasks, the average time for completion was reduced by 32.5\% when using \systemname (Mean = 292.1 seconds, $\sigma$ = 56.9 seconds) compared to using \unakite (Mean = 432.7 seconds, $\sigma$ = 61.8 seconds), which is also statistically significantly (p < 0.05). Thus, using \systemname did help participants evaluate the appropriateness for reuse faster.

\begin{figure}[t]
  \centering
  \begin{subfigure}{.47\textwidth}
  \centering
  \includegraphics[width=0.95\linewidth]{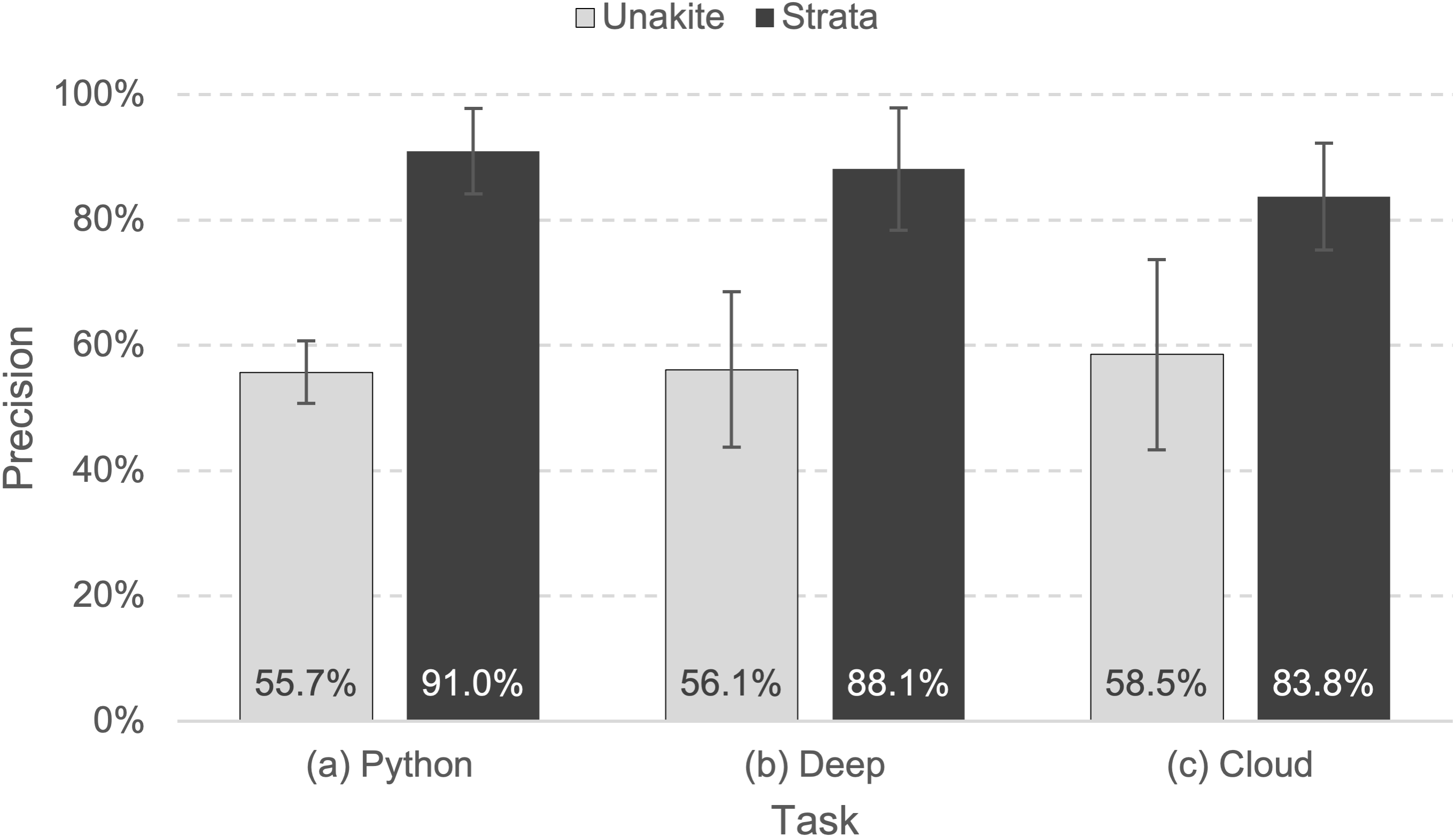}
  \caption{Precision of high quality reasons ($n_\textbf{High Quality} / n_\textbf{Total}$.)}
  \label{fig:sub1}
\end{subfigure}\hspace{5mm} %
\begin{subfigure}{.47\textwidth}
  \centering
  \includegraphics[width=0.95\linewidth]{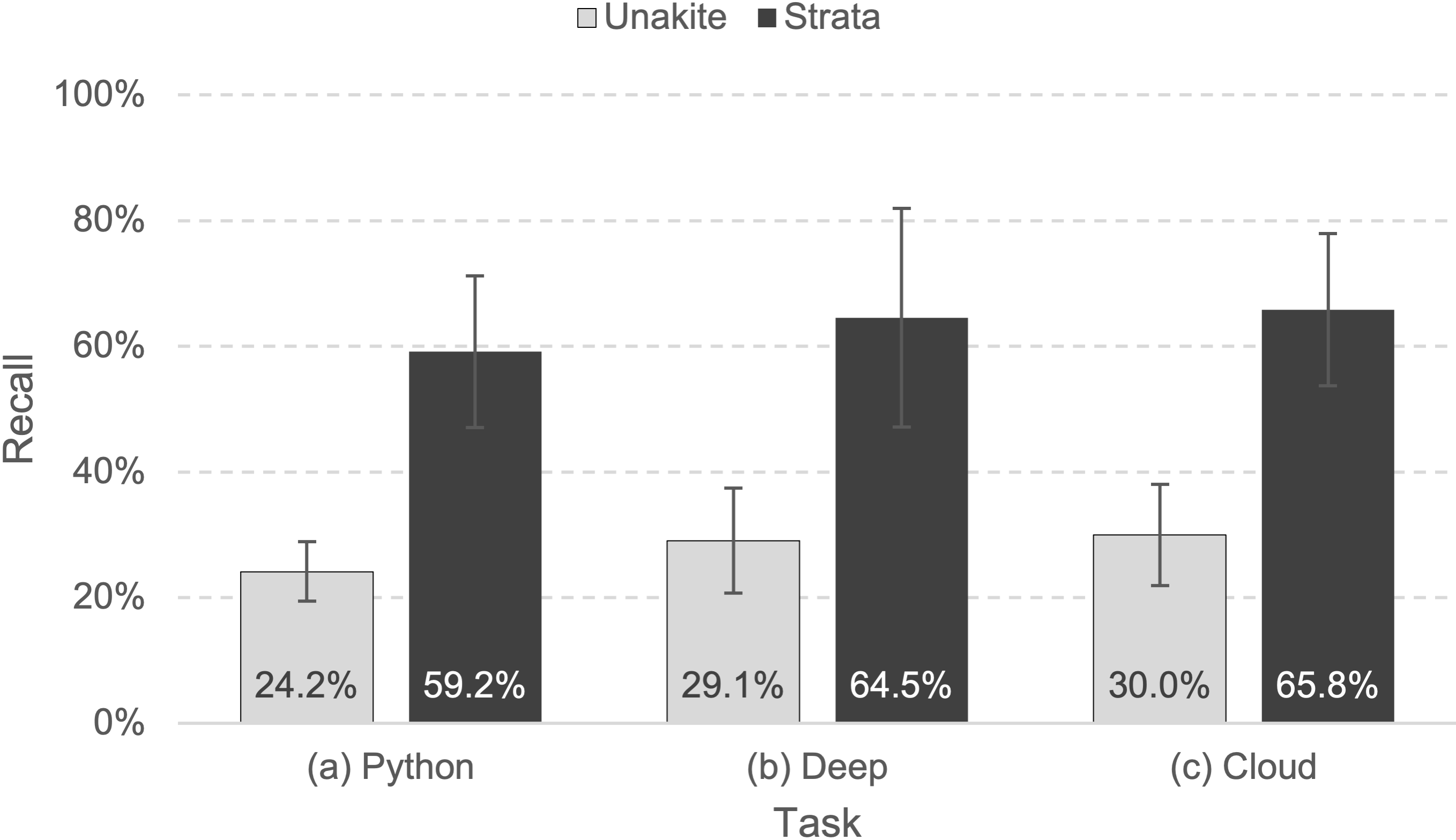}
  \caption{Recall of high quality reasons ($n_\textbf{High Quality} / n_\textbf{Ref. High Quality}$.)}
  \label{fig:sub2}
\end{subfigure}
\caption{Precisions and recalls of high quality answers in all three tasks. All results are statistically significant under t-tests (p < 0.05).}
\label{fig:precision-recall}
\end{figure}

To assess the \textit{quality} of the reasons that participants came up with, before the study,  two professional developers who are not affiliated with the research each generated a list of \textit{high quality} reasons for all three tables independently. After resolving conflicts through discussions between the two developers, we produced a list of high-quality reasons for each table as the ``gold standard''. We then calculated and report in Table \ref{tab:study-results-table} the numbers of high quality reasons participants identified that are on the ``gold standard'' list, as well as the precision (calculated as $n_\textbf{High Quality} / n_\textbf{Total}$) and recall (calculated as $n_\textbf{High Quality} / n_\textbf{Ref. High Quality}$) of high-quality reasons (where $n_\textbf{Total}$ is the total number of reasons they generated, and $n_\textbf{Ref. High Quality}$ is the number of ``gold standard'' high-quality reasons for each task). By plotting the precisions and recalls in Figure \ref{fig:precision-recall}, we can see that participants in the \systemname condition achieved higher precision in all three tasks, that is, they gave a higher percentage of high-quality reasons in their responses compared to the \unakite condition. Participants in the \systemname condition also achieved higher recall in all three tasks, that is, they were able to find more high-quality reasons compared to the \unakite condition. Thus, using \systemname did help participants improve the quality of their evaluations compared to using \unakite.

In case participants came up with valid answers we had not thought of, after the study, we asked the same two developers as above to rate each reason that participants gave as either \textit{valid} or \textit{not valid} blind to the conditions. Valid reasons are considered as the ones that are specific and correct according to the content of the table. After resolving conflicts through discussions between the two developers, we filtered out the reasons that are considered \textit{invalid}, and presented the resulting numbers of valid reasons in Table \ref{tab:study-results-table} (the numbers of invalid reasons were negligible and were therefore not included in the table). Across all three tasks, the average total number of valid reasons ($n_\textbf{Valid}$) increased by 75.6\% when using \systemname (Mean = 7.90, $\sigma$ = 1.90) compared to using \unakite (Mean = 4.50, $\sigma$ = 0.94), which is also statistically significant (p < 0.05). Thus, using Strata appeared to help participants come up with more valid evaluations for appropriateness for reuse compared to \unakite alone.

In the survey, participants reported (in 7-point Likert scales) that they thought the interactions with \systemname were understandable and clear (Mean = 6.20, Median = 6.00, 95\% CIs = [5.75, 6.46]), they enjoyed \systemname's features (Mean = 6.00, Median = 6.00, 95\% CIs = [5.45, 6.72]), and would recommend \systemname to friends and colleagues (Mean = 6.10, Median = 6.00, 95\% CIs = [5.65, 6.35]).

\subsection{Qualitative Results}
\subsubsection{Usability and usefulness of \systemname's features}
Overall, participants appreciated the increased transparency and efficiency afforded by various \systemname features and highlighted the values of the appropriateness properties that we visualize, arguing that \textit{``it helps me understand how a table was made step by step''} (P10), \textit{``lets me know what the author searched for, so if I don't understand something, I can search again. And more importantly, I can sort of know what the author didn't look for, and sometimes that'll become what I can do next''} (P4), \textit{``[the automatic context snapshot feature] saves me lots of time that I would otherwise spend going to the source web pages and making sense of things, which could be a rabbit hole sometimes''} (P15), and \textit{``[allows me to] see on a high-level where stuff comes from and if there's any source that is potentially questionable''} (P13). 
\revision{In addition, P8 reflected that \systemname \textit{``serve(d) as a guidance for things that I should pay attention to,''} which underlines the value of our framework, and reminded some participants of appropriateness properties that they would otherwise overlook, such as \textit{``I never really thought about what the author(s) looked for or not, but now I think it's actually quite important, especially if they miss obvious things that an expert would never miss,''} (P6) and \textit{``I realize that I'm more of a grab-and-go kinda person and I don't usually remember to check how many up-votes a \stackoverflow answer gets or when it was last updated''} (P17).}


\subsubsection{Authoring tables}
Participants were also excited about authoring tables with \systemname running, as it will automatically extract and produce the sidebar on the left and the various signals in the table. They mentioned that such \textit{``honest signals enhanced''} (P10) tables would be particularly useful in situations such as code reviews (P6: \textit{``going through the three main aspects is like going through our usual quality checklist, which makes sure that we're not missing anything''}) and project takeovers (P13: \textit{``if my previous browsing sessions are captured by this, then I won't need to make myself available again and again if somebody else suddenly has a question that only I know the answer to, since I made it in the first place---this table thing will almost be self-explanatory''}).

\subsubsection{Privacy concerns}
Some participants shared their privacy concerns from an author's perspective, mentioning that certain types of metadata that could reveal their personal preferences and idiosyncrasies (e.g., the code that they used, the snippet surroundings, and their search queries) should be kept private until they felt comfortable sharing. Indeed, prior work has pointed out that there may be negative effects of surfacing certain types of information \cite{erickson_social_2000}. These findings identified new research opportunities for \revision{(1) intelligent mechanisms that can automatically screen for and block out information that should be kept private (e.g., similar to \cite{li_privacy-preserving_2020} or \cite{jin_why_2018}) and (2) mixed-initiative and interactive mechanisms \cite{horvitz_principles_1999} that collaborate with users to only preserve the information that they are comfortable sharing (e.g., similar to \cite{li_multi-modal_2020}) without compromising the usability and effectiveness of the system}.

%
%
%
%
%
%
\section{Discussion}
\label{sec:discussion}
\revision{
Prior research on web credibility stressed the importance of trustworthiness measurement during the evaluation of the appropriateness to reuse a previously created knowledge artifact \cite{hoorn_web_2010}. However, as we found from literature on sensemaking handoff and our formative study, evaluating the appropriateness of reuse is much more than simply verifying the trustworthiness \cite{hoorn_web_2010}, especially since the artifacts are often an author's collection and synthesis of different individual pieces of information from different sources and reflect the author's opinion about the trade-offs among multiple valid options \cite{liu_unakite:_2019}. As a result, in addition to understanding whether the content is trustworthy, consumers also need to understand if the original problem context when the author created the artifact matches with the consumer's \cite{markus_toward_2001,hoorn_web_2010}, and if the author's research process was thorough \cite{paul_cosense:_2009,dourish_awareness_1992}. One of the contributions that we make in this work is a framework (Table \ref{tab:framework-table}) that summarizes the aforementioned three major facets, serving as a checklist that guides consumers through their evaluation processes. \systemname, which is an instantiation of the framework, improves consumers' abilities to evaluate these facets compared to using \unakite alone, as evidenced by both the quantitative (i.e., number of valid reasons given by the participants in terms of each facet) and qualitative  results (e.g., participants' comments on \systemname reminding them of double checking appropriateness properties that they would otherwise overlook).
}

\revision{
Although prior work on trust and sensemaking handoff offers insights into the various aspects and properties that are important for evaluating the appropriateness of reuse, it remained costly and difficult for not only the author who was creating the knowledge to also keep track of those signals and save them somewhere (since it is extra work without immediate benefit), but also for the consumer who was interpreting the knowledge to deduce and speculate about those signals. Through our research, we learned that an reasonable number of appropriateness signals can automatically be captured at authoring time as well as processed and visualized to the consumers subsequently to help with the reuse evaluation, and thereby reduce the cost for people to build on each other's knowledge artifacts.
}


\section{Future Work}

One participant (P4) in the evaluation study said \textit{``I can imagine myself having this table page open as I collect stuff so I can check how well I'm doing as I go''}, which suggests that \systemname not only can help consumers but also provides value for authors at collection time --- authors can use \systemname features to help them know how well their decisions will be judged, how thorough they have been, whether they are using up-to-date materials, if there have been any version mismatches, etc. In the future, we would like to \revision{investigate how to} integrate these \systemname visualization features into authors' workflows to help them ``proofread'' their decision making processes in real time.

Currently, \systemname has settings that consumers can tune based on their personal preferences, such as the trusted domain whitelist. Future work is needed to investigate mechanisms that can enable consumers to also personalize an existing table, such as adding, editing, and removing certain elements, effectively creating new versions of that table without overriding the original author's version. In addition, it would also be an interesting challenge to aggregate the changes in different consumers' versions and propagate them back to the original author as constructive feedback. 


Finally, our approach may have potential implications for other situations and domains involving user-generated content (beyond comparison tables), in which knowledge consumers need to evaluate the relevance and trustworthiness of that content. For example, the context, trustworthiness, and thoroughness facets could provide generative inspirations for helping users evaluate how knowledge artifacts were constructed, such as in Wikipedia (e.g., which sources were considered for an article, properties of the contributors, and coverage of key topics mined from similar articles), Q\&A sites like \stackoverflow where many people collaborate and edit questions and answers together;  curation platforms such as Pinterest, or thousands of other wiki systems. Generalizing how to augment knowledge reuse for situations beyond decision making in programming is an interesting and potentially fruitful area for future investigation, including exploring which information needs identified in this paper may not be as relevant and which additional needs become important. On the one hand, such an endeavor could unlock cycles of knowledge reuse in which people can quickly make good judgements about which information to aggregate and accumulate, which then become useful signals for making future judgements easier as well. On the other hand, the various signals and properties that are automatically surfaced could raise consumers' awareness of the potential existence of mis-information online \cite{vicario_spreading_2016} and provide readily available evidence to combat it.

%
%
%
%
%
%
\section{Limitations and Risks}
\label{sec:limitations}

There are certain types of information that \systemname is not able to automatically obtain and visualize. One set of limitations results from \systemname working in the browser, so it cannot monitor activities which happen in the authors' code editors or IDEs, \revision{command line interfaces}, and relevant discussions with friends and colleagues (\revision{communicated} either verbally or electronically through chat applications like Slack). Further development of extensions in these different environments as well as research into how to coordinate the collection and organization of this information would be needed in order to provide consumers with a more complete picture of an authors' working context beyond the browser. \revision{However, even in situations where \systemname cannot automatically calculate a signal, we believe that the three major facets still alert consumers that these are important aspects to be considered. Also, to the extent that consumers come up with their own measurements and ways to fulfill their information needs, they are perfectly welcome to do so, such as testing if a piece of sample code returns the desired result by running it in a terminal, which the current \systemname does not automatically do.}

Some of the features in \systemname are currently implemented based on heuristics, such as the bounds of the automatic context snapshots and the threshold beyond which information is considered out-of-date. These heuristics are based on our preliminary piloting through limited iterations, and may not apply universally to every situation. Further development can make these features more universally applicable and more adaptive to different situations so that users will be able to rely more on the judgements that \systemname automatically generates.

The current design of \systemname is intended for use cases where people collaborate and communicate their knowledge artifacts with each other in good faith; for example, software engineers sharing design rationale within a team. However, for \systemname to be used at scale with potentially malicious actors, such as in situations where some authors might try to increase the trustworthiness and thoroughness scores by manipulating the different metrics that it uses and displays, additional signals as well as mitigation techniques might be needed to combat such gaming behaviors. One approach would be to aggregate multiple knowledge artifacts with similar context (options, criteria, and goals in the case of \unakite comparison tables) together and detect and filter out anomalous components, inspired by mechanisms like ``down-voting'' that community Q\&A sites (e.g., \stackoverflow) use to guard against incorrect and malicious answers at scale. Further, some of the information, like the context, seems difficult and pointless to distort.


One of the concerns that repeated during our iterative design process is that each surfaced appropriateness property ultimately competes for user attention and takes time for the reader to process \cite{kittur_can_2008}, which could result in the overall user interface being overwhelming. The current solution we employed, inspired by prior work in recursive summarization and sensemaking \cite{zhang_wikum:_2017,zhang_making_2018}, takes a hierarchical approach that presents users with an overview and the ability to dive into specific details, letting them take the initiative of exploring parts relevant to their own interests. Future research is needed to untangle the relative importance of the various factors and how they can be alternatively represented. One idea is to gather large amounts of usage data from a field deployment and develop statistical or machine learning-based models that can predict importance metrics given various input parameters.

Finally, our lab evaluation contains several limitations. Given the short amount of training time participants had, some may not have been able to get fully acquainted with the various features that \systemname offers. The tasks used in the study may not be what participants encounter in their daily work, and participants may not have the necessary context and sufficient agency as they do in real life. We mitigated these risks by asking participants to complete a practice task simulating what they would need to do in the study to help them get familiarized with \systemname as well as the flow and cadence of the tasks. To improve realism, all three tasks used in the study were based on actual questions asked by real developers online, and the tables used in the study were adapted by the first author from real comparison tables we found online. For each task, we also provided participants with some background information and context to get them prepared. In the future, we would like to further address these limitations by conducting a long-term larger-scale field study, where developers will have both sufficient familiarity with \systemname through repeated usage and motivation to reuse decisions that are relevant to their own work.

%
%
%
%
%
%
\revision{
\section{Conclusion}
\label{sec:conclusion}
Appropriate reuse of previously created knowledge requires judging its relevance, trustworthiness, and thoroughness in relation to an individual's goals and context. In this work, we synthesized a framework for such reuse judgements in the domain of programming through analysis of prior research on sensemaking and trust as well as new needs-finding interviews with developers. In addition, we developed a prototype system called \systemname that automatically captures and visualizes some of the signals described in the framework that would facilitate subsequent knowledge consumers' reuse decisions, which proved to be effective and useful in a user study.

\unakite and \systemname together point to the importance of having tool support that helps people more efficiently organize and manage information as they find it in a way that could also be beneficial to others, and therefore bootstrapping the virtuous cycle of people being able to build on each other's sensemaking results, fostering efficient collaboration and knowledge reuse.
}

%
%
%
%
%
%
\begin{acks}
  This research was supported in part by NSF grants CCF-1814826 and FW-HTF-RL-1928631, Google, Bosch, the Office of Naval Research, and the CMU Center for Knowledge Acceleration. Any opinions, findings, conclusions, or recommendations expressed in this material are those of the authors and do not necessarily reflect the views of the sponsors. We would like to thank our study participants for their kind participation and our anonymous reviewers for their insightful feedback. We are grateful to Amber Horvath, Toby Jia-Jun Li, Haojian Jin,  Joseph Chee Chang, Nathan Hahn, Zheng Yao, Yiyi Wang, Tianying Chen, Haitian Sun, Jiachen Wang, and Jinlei Chen for their valuable feedback and constant support, especially during the COVID-19 pandemic.
\end{acks}

\bibliographystyle{ACM-Reference-Format}
\bibliography{aaareferences}

\end{document}